\documentclass[ALICE,manyauthors]{cernphprep}
\begin{document}%
%
%
%
\newcommand {\pT}{\ensuremath{p_{\mathrm{T}}}}
\newcommand {\meanpT}{$\langle \pT \rangle$}
\newcommand {\vf}{\ensuremath{v_{\mathrm{2}}}}
\newcommand {\et}{\ensuremath{E_{\mathrm{t}}}}
\newcommand {\mT}{\ensuremath{m_{\mathrm{t}}}}
\newcommand {\ee}{\mbox{e$^+$e$^-$}}
\newcommand {\modrap} {$\left | y \right | $}
\newcommand {\sigee}{$\sigma_E$/$E$}
\newcommand {\dndy}{d$N$/d$y$}
\newcommand {\dndydpT}{d$^{\mathrm 2}N$/d$y$d\pT}
\newcommand {\dNdpt}{d$N$/d\pT}
\newcommand {\dnchdy}{d$N_{\mathrm{ch}}$/d$y$}
\newcommand {\dedx}{d$E$/d$x$}
\newcommand {\PbPb}{\mbox{Pb--Pb} }
\newcommand {\pp}{pp}
\newcommand {\s}{$\sqrt{s}$}
\newcommand{\ppbar} {\mbox{$\mathrm {p\overline{p}}$}}
\newcommand {\lum} {\, \mbox{${\rm cm}^{-2} {\rm s}^{-1}$}}
\newcommand {\barn} {\, \mbox{${\rm barn}$}}
\newcommand {\m} {\, \mbox{${\rm m}$}}
\newcommand {\dg}{\mbox{$^\circ$}}
\newcommand {\stat}{({\it stat.})~}
\newcommand {\syst}{({\it syst.})~}
\newcommand {\degree}{$^{\rm o}$}
%
%
\newcommand {\tev} {TeV}
\newcommand {\gev} {GeV}
\newcommand {\mev} {MeV}
\newcommand {\kev} {keV}
\newcommand {\gmom} {GeV/$c$}
\newcommand {\mmom} {MeV/$c$}
\newcommand {\mmass} {MeV/$c^{\mathrm 2}$}
\newcommand {\gmass} {GeV/$c^{\mathrm 2}$}
\newcommand {\nb} {\mbox{\rm nb}}
\newcommand {\musec} {$\mu$s}
\newcommand {\nsec} {ns}
\newcommand {\psec} {ps}
\newcommand {\fm} {fm}
\newcommand {\cm} {cm}
\newcommand {\mm} {mm}
\newcommand {\mim} {$\mu$m}
\newcommand {\cmq} {cm$^{\mathrm 2}$}
%
%
\newcommand{\pizero}{\mbox{$\mathrm {\pi^0}$}}
\newcommand{\K}{\mbox{$\mathrm {K}$}}
\newcommand{\Kzs}{\mbox{$\mathrm {K^0_S}$}}
\newcommand{\rmLambda}{\mbox{$\mathrm {\Lambda}$}}
\newcommand{\rmAlambda}{\mbox{$\mathrm {\overline{\Lambda}}$}}
\newcommand{\rmXi}{\mbox{$\mathrm {\Xi^{-}}$}}
\newcommand{\rmAxi}{\mbox{$\mathrm {\overline{\Xi}^{+}}$}}
\newcommand{\rmOmega}{\mbox{$\mathrm {\Omega}$}}
\newcommand{\Xis}{\mbox{$\mathrm {\Xi^{-}+\overline{\Xi}^{+}}$}}
\newcommand{\kstar}{\mbox{K$^{*}$(892)$^{\mathrm {0}}$}}
\newcommand{\phir}{\mbox{$\mathrm {\phi}$(1020)}}
\newcommand{\rmphi}{\mbox{$\mathrm{\phi}$}}
\newcommand{\Sigmastar}{\mbox{$\mathrm {\Sigma}$(1385)$^{\pm}$}}
\newcommand{\Xistar}{\mbox{$\mathrm {\Xi}$(1530)$^0$}}
\newcommand{\Jpsi} {\mbox{J\kern-0.05em /\kern-0.05em$\psi$}}
\newcommand{\psip} {\mbox{$\psi^\prime$}}
\newcommand{\Ups} {\mbox{$\Upsilon$}}
\newcommand{\Upsp} {\mbox{$\Upsilon^\prime$}}
\newcommand{\Upspp} {\mbox{$\Upsilon^{\prime\prime}$}}
\newcommand{\ccbar} {\mbox{$\mathrm {c\overline{c}}$}}
\newcommand{\bbbar} {\mbox{$\mathrm {b\overline{b}}$}}
\newcommand{\ssbar} {{s$\overline{\mathrm s}$}}
\newcommand{\qqbar} {{q$\overline{\mathrm q}$}}
\newcommand{\qqqqbar} {{qq$\overline{\mathrm {qq}}$}}
\newcommand{\DDbar} {\mbox{$\mathrm {D\overline{D}}$}}
\newcommand{\ie}{i.e.\@\xspace}
%
%
\hyphenation{stran-ge-ness}
\hyphenation{pa-ra-me-tri-zed}
\hyphenation{PY-THIA}
\hyphenation{si-mu-la-tions}
\hyphenation{sta-ti-sti-cal-ther-mal}
\hyphenation{ther-mal}
%
\begin{titlepage}
\PHnumber{2012-221}                 
\PHdate{27 Aug 2012}              
%
%
\title{Production of \kstar~and \phir~in \pp~collisions at \s=7~\tev}
\ShortTitle{Production of \kstar~and \phir~in \pp~collisions at \s=7~\tev}   
%
\Collaboration{ALICE Collaboration%
         \thanks{See Appendix~\ref{app:collab} for the list of collaboration
                      members}}
\ShortAuthor{ALICE Collaboration}      
\begin{abstract}
The production of \kstar~and \phir~in \pp~collisions at \s=7~\tev~was measured by the ALICE experiment at the LHC.
The yields and the transverse momentum spectra {\dndydpT} at midrapidity \modrap $<$ 0.5 in the range 0$<$\pT$<$6~\gmom~for \kstar~and 0.4$<$\pT$<$6~\gmom~for \phir~are reported and compared to model predictions.
Using the yield of pions, kaons, and $\rmOmega$ baryons measured previously by ALICE at \s=7~TeV,
the ratios K$^*$/K$^-$, $\rmphi$/K$^*$, $\rmphi$/K$^-$, $\rmphi$/$\pi^-$, and $(\rmOmega + \overline{\rmOmega})/\rmphi$ are presented. 
The values of the K$^*$/K$^-$, $\rmphi$/K$^*$ and $\rmphi$/K$^-$ ratios are similar to those found at lower centre-of-mass energies.
In contrast, the $\rmphi$/$\pi^-$ ratio, which has been observed to increase with energy,
seems to saturate above 200~GeV. The $(\rmOmega + \overline{\rmOmega})/\rmphi$ ratio in the 
\pT~range 1-5 {\gmom}  
is found to be in good agreement with the prediction of the HIJING/B$\overline{\rm B}$ v2.0 model with a strong colour field.
\end{abstract}
\end{titlepage}
\setcounter{page}{2}
\section{Introduction}
\label{intro}
 
The study of resonance production plays an important role in both elementary and  heavy ion collisions.
In \pp~and e$^+$e$^-$ collisions it contributes to the understanding of hadron production~\cite{Aguillar:1991,Albrecht:1994} 
as the decay products of resonances represent a large fraction of the final state particles. In addition,  
it provides a reference for tuning  event generators inspired by Quantum Chromodynamics (QCD). 
In heavy ion collisions, resonances are a sensitive probe of the dynamical evolution of the fireball.
Due to their  short lifetime (a few fm/$c$) a significant fraction of resonances decay inside the hot and dense medium and their hadronic daughters 
 interact with the medium during the fireball expansion~\cite{Adams:2005,Adler:2005,Abelev:2009}.

The \phir, which is the lightest 
vector meson composed only of sea quarks, provides a probe for the study of the strangeness production. 
In \pp~collisions,  $s\bar{s}$ pair production was found to be significantly suppressed in comparison to $u\bar{u}$ 
and $d\bar{d}$-pair~\cite{Malhotra,Wroblewski}. 
Another useful probe of strangeness production is the \kstar, which is a vector meson with a mass similar 
to the $\rmphi$, but differing by one unit of the strangeness quantum number. 
The $(\rmOmega + \overline{\rmOmega})/\rmphi$ ratio has been suggested~\cite{Toporpop:2011} as a probe of the colour 
field strength, which in microscopic models influences the relative yield of strange with respect to non-strange particles.
 
We present the first measurement of the differential (\dndydpT) and \pT-integrated (\dndy) yields 
of the K$^*$~\footnote{We denote by K$^*$ the average of K$^*$(892)$^0$ and $\overline{{\rm K}^{*}{\rm(892)}^0}$.} and 
\phir~mesons at midrapidity (\modrap $<$0.5) in \pp~collisions at \s=7~TeV.
The data analysis was carried out for K$^*$ ($\rmphi$) on a sample of 80 (60) million minimum bias 
\pp~collisions collected by the 
ALICE experiment.
The resonances were identified via their main decay channel 
K$^* \longrightarrow \pi^{\pm}$+K$^{\mp}$ and $\rmphi \longrightarrow$ K$^{+}$+K$^{-}$.
Tracks were reconstructed by the main ALICE tracking devices, the Time Projection Chamber (TPC) and 
the Inner Tracking System (ITS). 
The TPC and Time of Flight (TOF) detectors were used to identify pions and kaons.
 The measured spectra are compared to two QCD-based event generators, 
PHOJET~\cite{PHOJET} and PYTHIA~\cite{PYTHIA}.

The ratios  K$^*$/K$^-$, $\rmphi$/K$^*$, $\rmphi$/K$^-$, and $\rmphi$/$\pi^-$  are computed 
using the yield of pions and kaons measured~\cite{Spectra_7} with the ALICE detector in \pp~collisions at 7~TeV. These ratios are
compared with measurements at lower collision energies.
The ($\rmOmega + \overline{\rmOmega})/\rmphi$ ratio has been calculated as a function of transverse momentum using 
the $\rmOmega$ and $\overline{\rmOmega}$ yield measured at 7~TeV~\cite{Omega_ALICE}; this ratio is then compared 
to the predictions of the HIJING/B$\overline{\rm B}$ v2.0 model with a Strong Colour Field (SCF)~\cite{LHCprediction} and to PYTHIA-Perugia~2011~\cite{Perugia_tunes}. 

The article is organized as follows: Section 2 gives details about the detectors relevant for this analysis,
Section 3 describes the criteria used for event selection, Section 4 gives an 
overview of the analysis, Section 5 presents the results and Section 6 the conclusions.

\section{Experimental set-up }
\label{sec:1}

A full description of the ALICE detector can be found in~\cite{ALICE_JINST,ALICE_all}.
For the analyses described in this paper, the ITS, the TPC, and the TOF detectors were used. 
These detectors are set inside a large solenoidal magnet providing a magnetic field $B$=0.5~T, and have a 
common pseudorapidity coverage of $|\eta|<0.9$.
Two forward scintillator hodoscopes (VZERO) placed along the beam direction at -0.9~m and 3.3~m on either side of the 
interaction point, cover the pseudorapidity regions $-3.7<\eta<-1.7$ and $2.8<\eta<5.1$. These are used for triggering 
and for rejecting  beam-gas interactions.

\subsection{The Inner Tracking System}
\label{sec:detits}
The ITS~\cite{ALICE_all} is the innermost ALICE detector, located between 
3.9 and 43~cm radial distance from the beam axis.
It is made of six cylindrical layers of silicon detectors (two layers of pixels, two of silicon drift, and two of 
silicon strips), with a total material budget of 7.66 \% of the radiation length $X_{\mathrm 0}$.
It provides high-resolution space points close to the interaction vertex, thus improving momentum and angular resolution 
of the tracks reconstructed in the TPC.

The two innermost ITS layers constitute the Silicon Pixel Detector (SPD), which has a high granularity of about 9.8 million 
pixel cells, each with a size of 50$\times$425~\mim$^{\rm 2}$. 
These layers are located at radii of 3.9 and 7.6~\cm~with pseudorapidity coverages of $|\eta|<2.0$ and $|\eta|<1.4$, 
respectively.
The detector provides a position resolution of 12~\mim~in the $r\varphi$ direction and about 100~\mim~in the direction 
along the beam axis. 

\subsection {The Time Projection Chamber}
\label{sec:dettpc}
The TPC~\cite{Alme:2010} is the main ALICE tracking device. It is a large-volume,  high-granularity,  cylindrical drift detector 
which has a length of 5.1~m and 
inner and outer radii of 0.85 and 2.47~m, respectively. 
It covers the pseudorapidity range $|\eta|<0.9$ with a full azimuthal acceptance.
The drift volume is filled  with 90~m$^{\rm 3}$ of Ne/CO$_{\rm 2}$/N$_2$.
The maximum drift time is 94~\musec.
A total of 72 multi-wire proportional chambers with cathode pad readout instrument the two end plates, 
which are segmented into 18 sectors and include a total of over 550,000 readout pads. 
The ionization electrons drift for up to 2.5~m and are measured on 159 pad rows. 
The momentum resolution of the TPC is in the range 1-7\% for pions with
1$<$\pT$<$10~\gmom. 
The ALICE TPC ReadOut (ALTRO) chip, employing a 10 bit ADC at 10~MHz sampling rate and digital filtering circuits, 
allows for precise position and linear energy loss measurements with a gas gain of the order of $10^4$.
The material budget of the TPC near $\eta=0$ amounts to about 4.1\% of $X_{\mathrm 0}$. 

The position resolution in the $r\varphi$ direction varies between 1100~\mim~and 800~{\mim} going from the inner to 
the outer radius, whereas the resolution along the beam axis varies between 1250~\mim~and 1100~\mim.
\subsection {The Time Of Flight detector}
\label{sec:dettof}

The ALICE TOF~\cite{Akindinov:2009,Akindinov:2010} is a cylindrical assembly of Multi-gap Resistive Plate Chambers (MRPC) 
with an inner radius 
of 370~cm and an outer radius of 399~cm. It has a pseudorapidity  coverage of
$|\eta|<0.9$ and full azimuthal acceptance, except for the region 260\degree $< \varphi <$ 320\degree~at 
$|\eta|<0.14$ where a gap was left in order to reduce the amount of material in front of the Photon Spectrometer (PHOS).
The elementary unit of the TOF system is a 10-gap double-stack MRPC strip 122~cm long and 13~cm wide, with an active 
area of 120$\times$7.4~\cm$^{\rm 2}$ subdivided into two rows of 48 pads of 3.5$\times$2.5~\cm$^{\rm 2}$ each. 
The length of the TOF barrel active region is 741~cm.
It has about 153,000  readout channels and an average thickness of 25-30\% of $X_{\mathrm 0}$, depending on the detector zone.
For \pp~collisions, such a segmentation leads to an occupancy below 0.02\%. 
The front-end electronics are designed to comply with the basic characteristics of the MRPC detector,  i.e. 
very fast differential signals from the anode and 
the cathode readout: the resulting intrinsic time
resolution of the detector and electronics was measured to be smaller than 50~ps.
\section{Data collection and event selection}
\label{sec:event}
Data used for this analysis were collected in 2010 using a magnetic field of $B$=0.5~T with both field polarities.
The minimum bias trigger required a single hit in the SPD detector or in one of the two VZERO counters, 
i.e. at least one charged particle anywhere in the $\sim$8 units of pseudorapidity covered by these detectors.
In addition, a coincidence was required with signals from two beam pick-up counters, 
one on each side of the 
interaction region, indicating the passage of proton bunches.
The trigger selection efficiency for inelastic collisions was estimated to be 
85.2\% with a +6.2\% and -3\% relative uncertainty~\cite{ALICE_cross_section}.
During the data-taking period, the luminosity at the ALICE interaction point was kept in the range 
$0.6-1.2\times 10^{\rm 29}$~cm$^{\rm {-2}}$s$^{\rm {-1}}$.
Runs with a mean pile-up probability per event larger than 5\% were excluded from the analysis.

Beam-induced background was reduced to a negligible level ($<0.01$\%) with the help of the timing information of the VZERO 
counters and by a cut on the position of the primary vertex reconstructed by the SPD~\cite{ALICE_first}.
Accepted events were required to have a reconstructed primary vertex.
Its position can be computed either using the tracks reconstructed by TPC and ITS, or using the ``tracklets'' obtained 
connecting reconstructed clusters in both SPD layers. If possible, the first method is used.
First, for each event a three dimensional 
reconstruction of the primary vertex was attempted with 
either a Kalman filter, using reconstructed tracks as input,
or by a minimization of the squared distances between all
the extrapolated tracklets.  Otherwise only the $z$ position of the primary vertex was  reconstructed by correlating the $z$ coordinates of 
the SPD space points, while for $x$ and $y$ the average position of the beam in the transverse plane was taken. 
The primary vertex reconstrunction efficiency, calculated via Monte Carlo simulation, 
approaches unity in events with a K$^*$ or a $\rmphi$ produced in the central rapidity region.
In order to minimize acceptance and efficiency biases for tracks at the edge of the TPC detection volume, 
events were accepted only when their primary vertex was within $\pm10$~cm from the geometrical centre of the ALICE barrel.
\section{Data analysis}
\label{sec:data}
\subsection{Track selection}
\label{sec:track}
Global tracking in ALICE is performed using ITS and TPC clusters.
It is based on a Kalman filter algorithm which takes into account both multiple scattering and  energy loss along 
the path as described in detail in~\cite{ALICE:PPR2}.
The Distance of Closest Approach (DCA) to the primary vertex is used to discriminate between primary and secondary particles.
Primary charged particles are those produced directly in the interaction and all decay products from particles with a proper 
decay length $c\tau<$1~cm; secondary particles include those from the weak decay of strange hadrons and from interactions in the detector material.
Several cuts were applied to achieve a high track quality in the analyzed sample.
Tracks were required to have at least 70 reconstructed clusters in the TPC out of the maximum 159 available.
This ensured a  high efficiency and good \dedx~resolution, keeping the contamination from secondary and fake tracks small.

In order to improve the global resolution, tracks were accepted only in the range $|\eta|<0.8$ (i.e. well within the 
TPC acceptance) and with \pT$> 0.15$~\gmom.
 In order to reduce secondary particles, tracks were required to have at least one hit in one of the two innermost 
 tracking detectors (SPD) and to have a DCA to the primary vertex less than 2~cm along the beam direction.  
 The DCA in the transverse plane was required to be smaller than 
7 $\sigma_{\mathrm {DCA}}$(\pT), where $\sigma_{\mathrm {DCA}}$(\pT) = (0.0026 + 0.0050~\gmom~$\cdot$\pT$^{-1}$)~cm. 
\subsection{Particle identification}
\label{sec:PID}
\begin{figure}
\resizebox{0.8\textwidth}{!}{
  \includegraphics{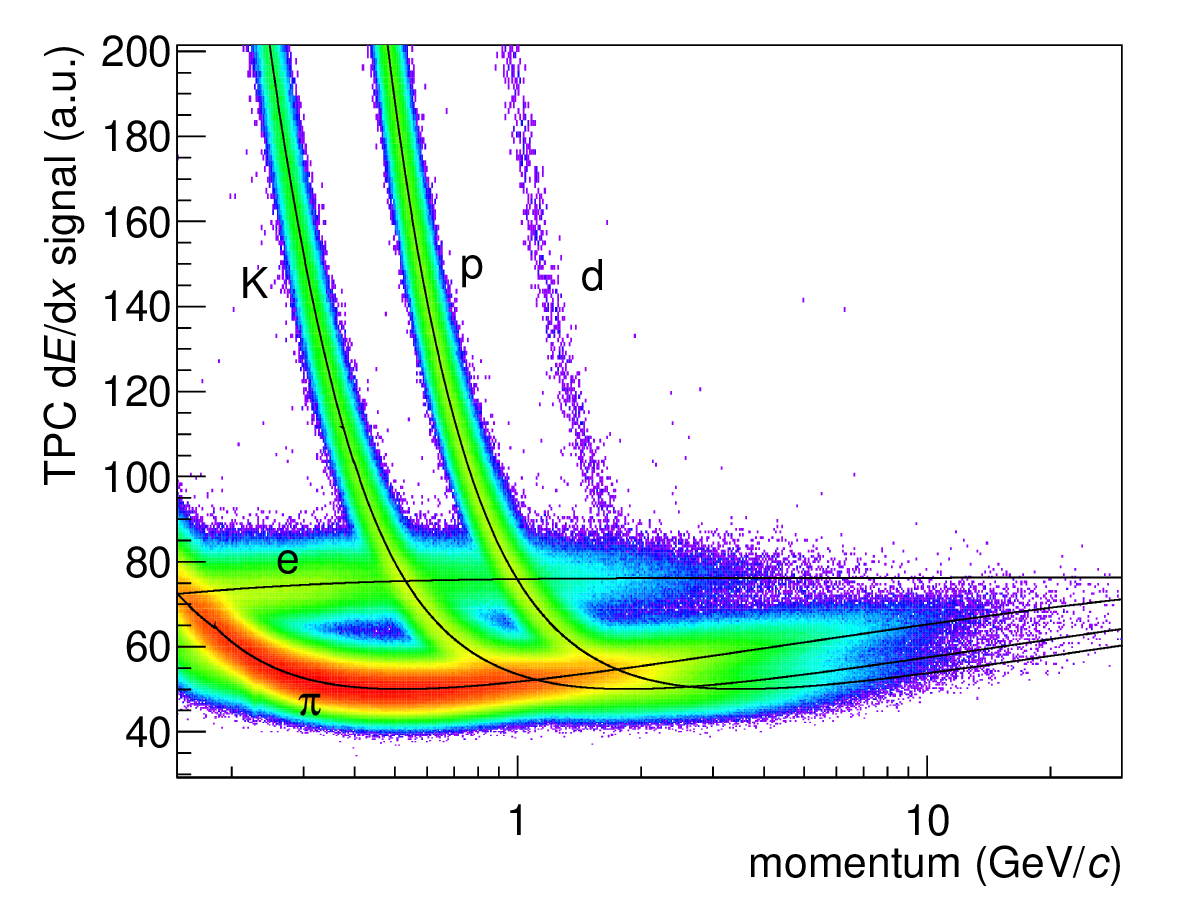}
}
\caption{(Colour online) Specific ionization energy loss \dedx~vs. momentum for tracks measured with the ALICE TPC. 
The solid lines are parametrizations of the Bethe-Bloch function~\cite{ALEPH}.}
\label{fig:TPCplot}
\end{figure}

Identification of pions and kaons is performed using the measurements of the TPC and the TOF.  
For the TPC, the particle  is identified based on the energy it deposits in the drift gas, compared with 
the expected value computed using a parameterized Bethe-Bloch function~\cite{ALEPH,Spectra_900}.
Figure~\ref{fig:TPCplot} shows the TPC signal versus track momentum computed at the 
point the particle enters the detector, 
and the curves represent the Bethe-Bloch functions for each mass hypothesis.
The TPC calibration parameters have mostly been determined and tested via the analysis of cosmic rays; 
 the chamber gain has been measured using the decay of radioactive $^{\rm 83}$Kr gas released into the TPC volume~\cite{Alme:2010}. 

A truncated-mean procedure is used to determine \dedx, with only 60\% of the points kept.
The \dedx~resolution $\sigma_{\mathrm {TPC}}$ is about 5\% for tracks with 159 clusters and about 6.5\% when
averaged over all reconstructed tracks. The relevant value of $\sigma_{\mathrm {TPC}}$ is estimated for each track 
taking into account the actual number of clusters used ~\cite{Alme:2010}.

The TPC \dedx~measurement allows   pions  to be separated 
from kaons for momenta up to $p \sim$ 0.7~\gmom, while the proton/antiproton band starts to overlap with the pion/kaon 
band at $p\approx$1~\gmom. 
As can be observed in Fig.~\ref{fig:TPCplot}, the electron/positron  \dedx~band crosses the other bands at various momenta.  
This contamination in identified pions and kaons can be drastically reduced using  information from the TOF.

Particles are identified in the TPC via the difference between the measured energy loss and the value expected for different mass hypotheses. 
The cut on this difference, normalized to the resolution $\sigma_{\rm {TPC}}$, is optimized for each analysis and depends 
in general on the signal-to-background ratio and on the transverse momentum.
\begin{figure}
\resizebox{0.8\textwidth}{!}{
  \includegraphics{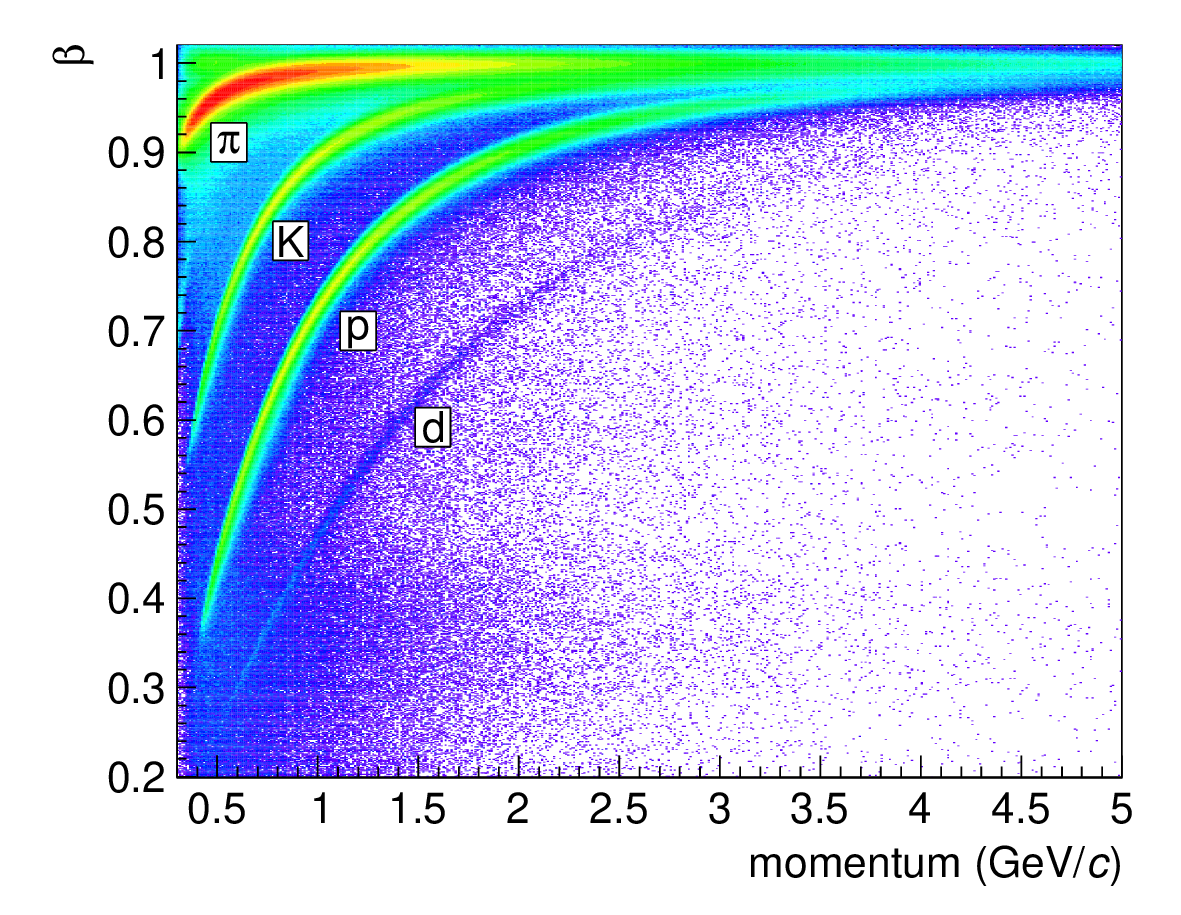}
}
\caption{(Colour online) Velocity $\beta$ of  particles measured by TOF vs. momentum.}
\label{fig:TOFplot}
\end{figure}
\begin{figure*}
\subfigure
{\resizebox{0.5\textwidth}{!}{
  \includegraphics{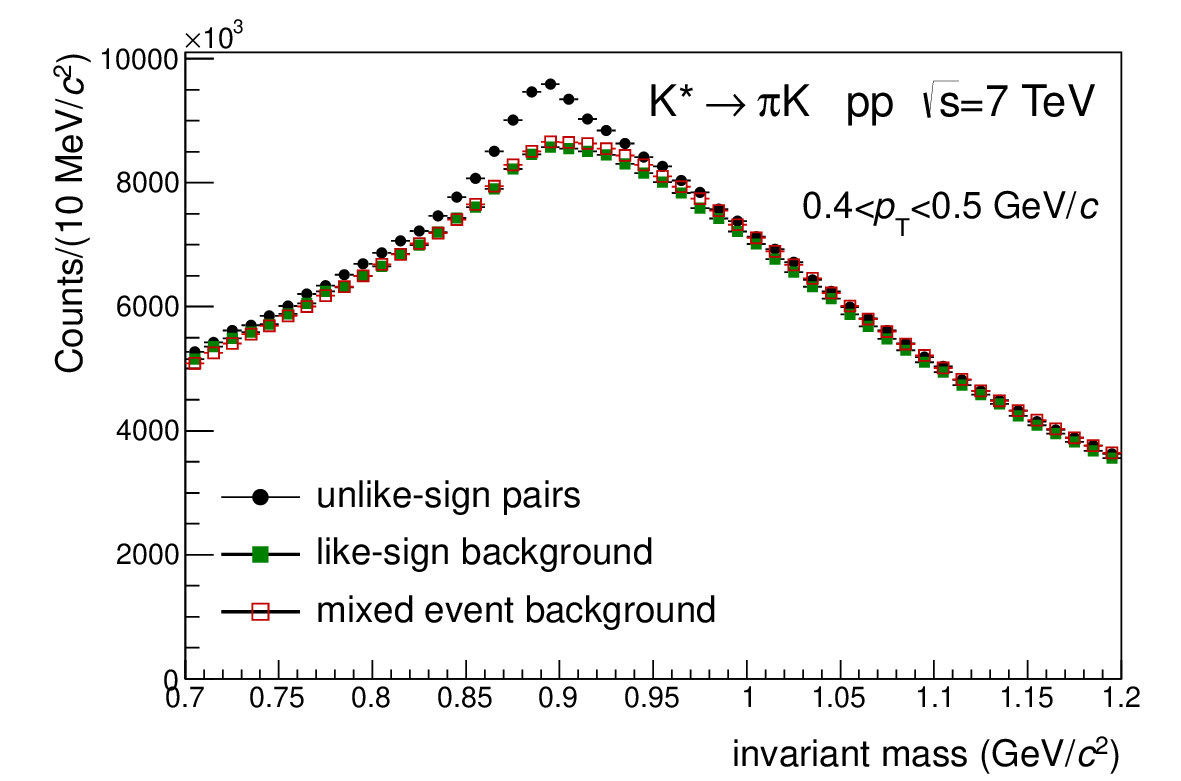}
}} 
\subfigure
{\resizebox{0.5\textwidth}{!}{
   \includegraphics{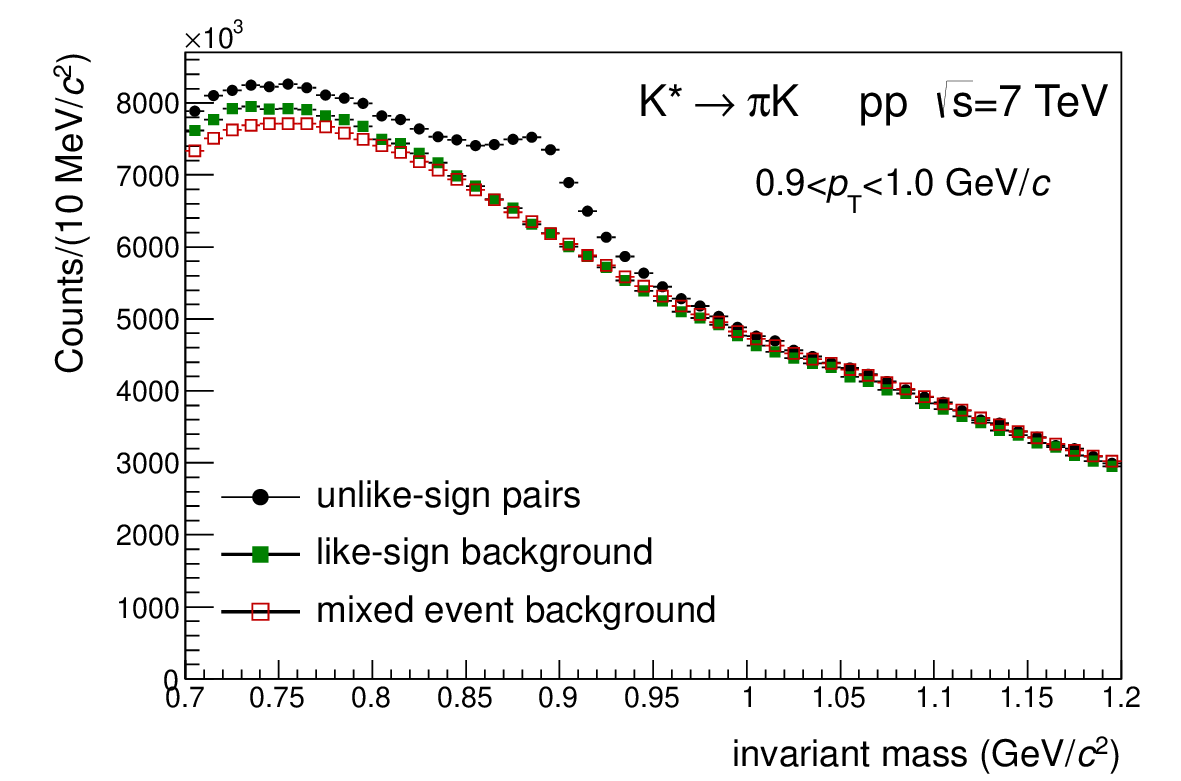}
}}

\subfigure {\resizebox{0.5\textwidth}{!}{
  \includegraphics{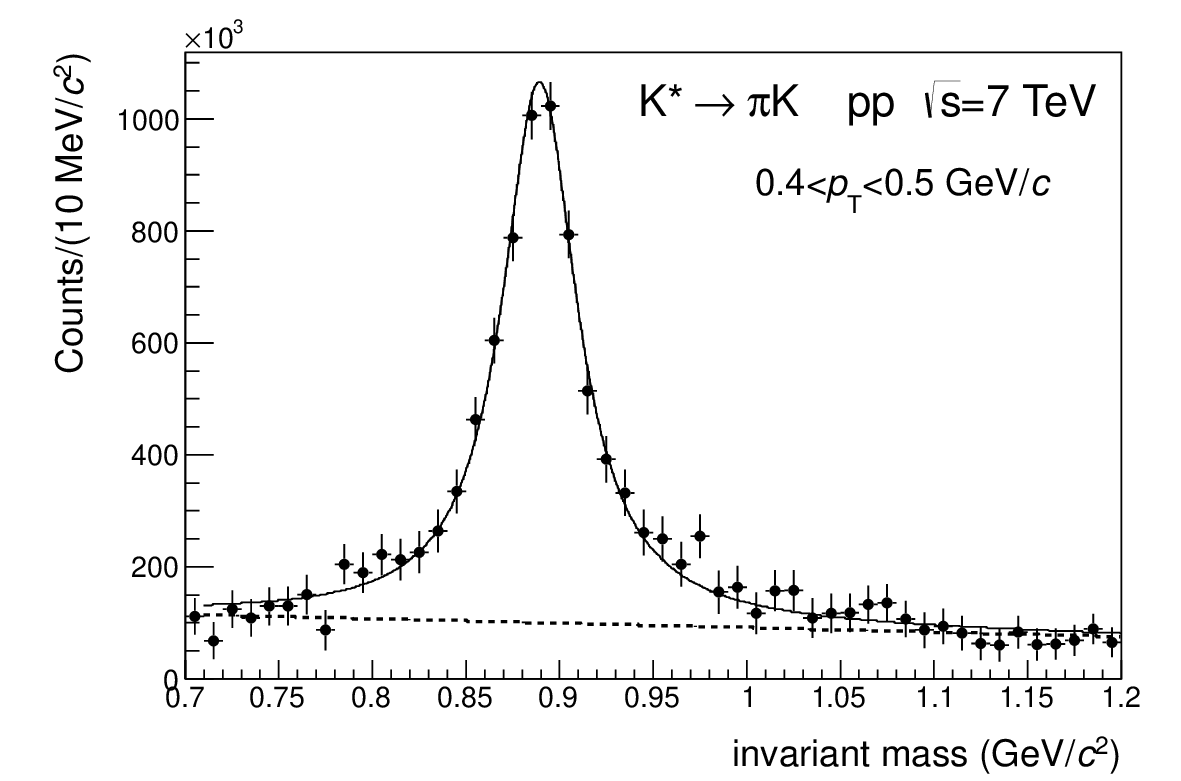}
}} 
\subfigure{
\resizebox{0.5\textwidth}{!}{
   \includegraphics{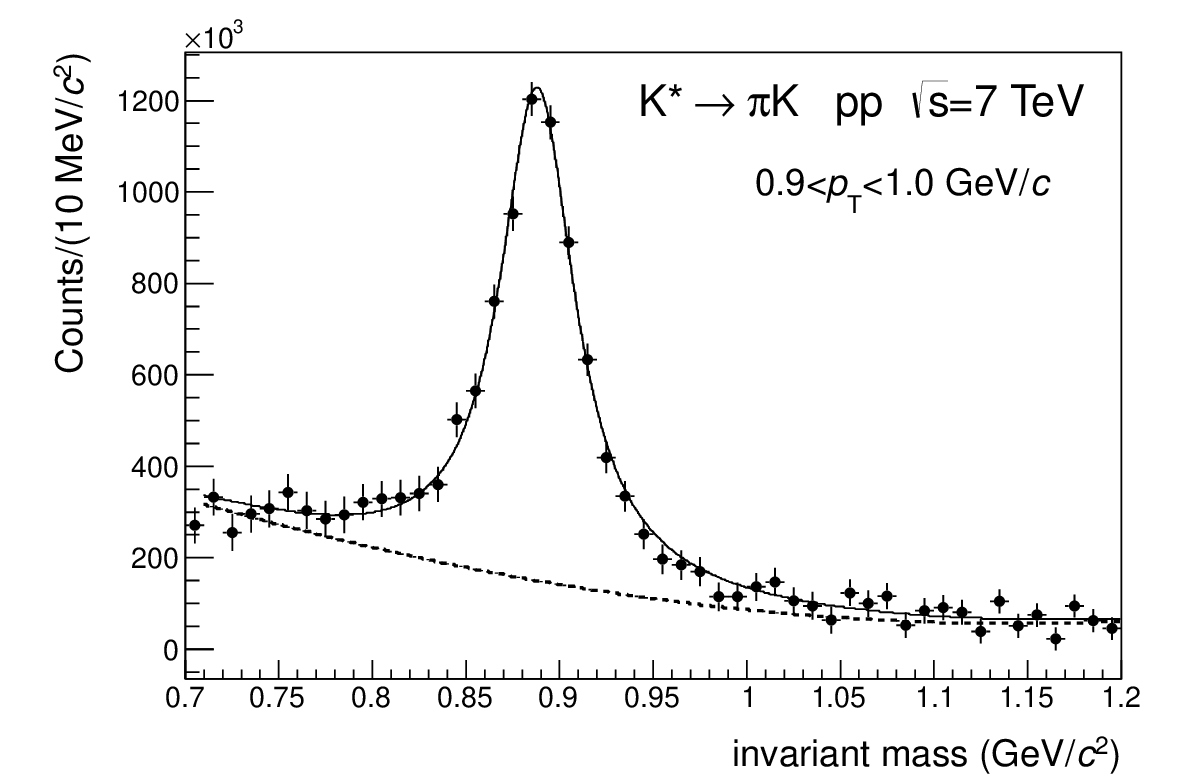}
}} 
   
   \caption{(Colour online) 
   (Upper panel) The $\pi^{\pm}$K$^{\mp}$  invariant mass distribution in \modrap$<$0.5 for the bin 0.4$<$
  \pT$<$0.5~\gmom~(left) and 0.9$<$\pT$<$1.0~\gmom~(right),  in \pp~collisions at 7~TeV.
   The background shape estimated using unlike-sign pairs from different events (event mixing) and like-sign pairs from the same event 
   are shown as open red squares and full green squares, respectively. 
   (Lower panel) The  $\pi^{\pm}$K$^{\mp}$ invariant mass distribution after like-sign background subtraction for 0.4$<$\pT$<$0.5~\gmom~(left) 
   and 0.9$<$\pT$<$1.0~\gmom~(right).    
   The solid curve is the result of the fit by Eq.~\ref{eqn:Breit-Wigner}, the dashed line describes the residual background.}
   \label{fig:invmass_Kstar}
\end{figure*}

Figure~\ref{fig:TOFplot} shows the correlation between particle momentum and their velocity $\beta = L/ct$, where $L$ is the total integrated path 
length and $t$ is the  time of flight measured by the TOF detector. 
For the analyses described in this paper the start time of the collision is estimated using the particle
arrival times at the TOF or the averaged collision time observed in the fill. The bands corresponding to pions, kaons, protons 
and deuterons are clearly visible.

Particles are identified in the TOF by comparing the  measured time of flight 
to the expected time for a given particle species. The cut
is expressed in units of the estimated  resolution $\sigma_{\mathrm {TOF}}$ for each track, which has a mean value of 160~ps.  
The TOF allows pions and kaons to be unambigously identifed up to $p\sim$ 1.5~\gmom.
The two mesons can be distinguished from (anti)protons up to $p\sim$ 2.5
\gmom.

\begin{figure*}
\subfigure {\resizebox{0.5\textwidth}{!}{
   \includegraphics{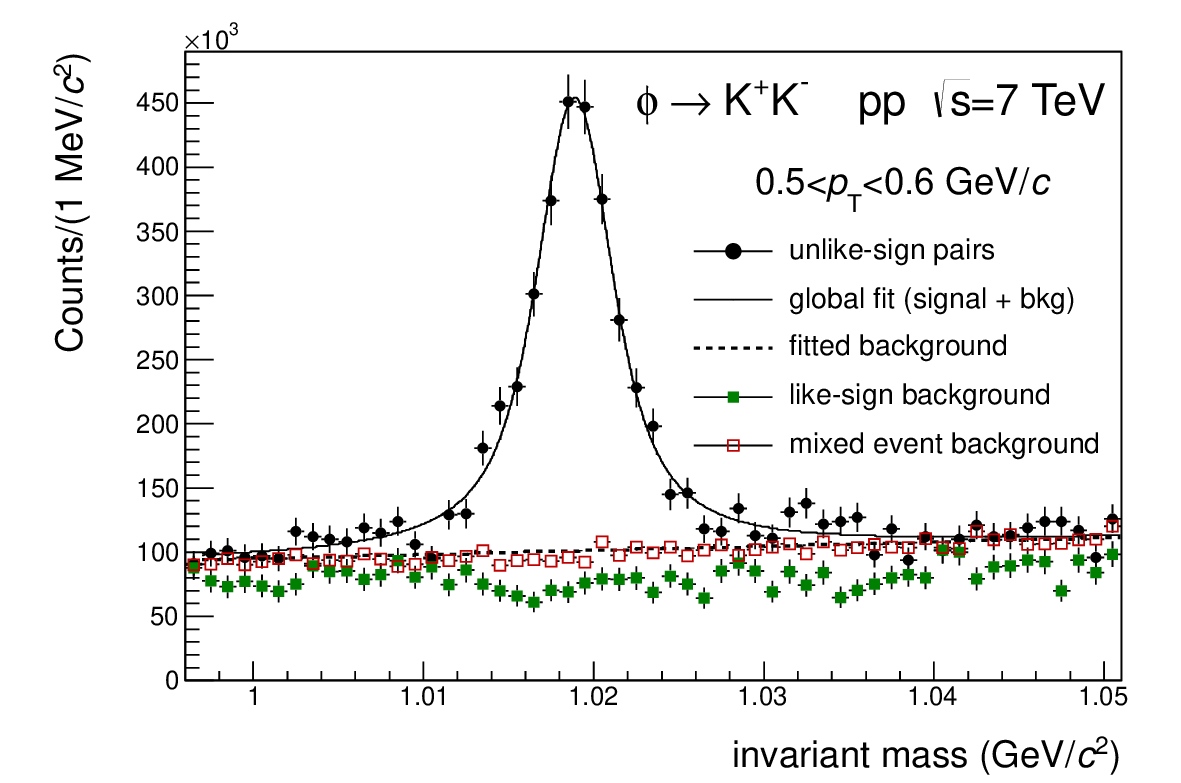}
}} 
\subfigure {\resizebox{0.5\textwidth}{!}{
   \includegraphics{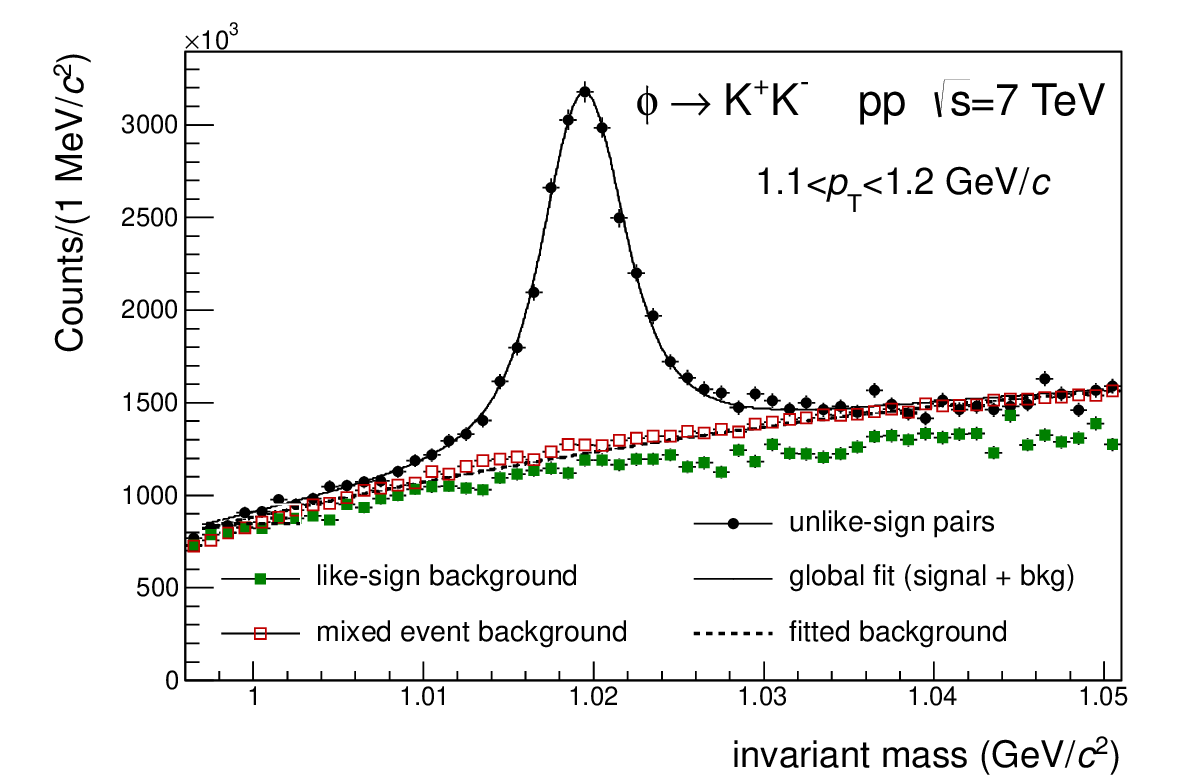}
}}

\subfigure {\resizebox{0.5\textwidth}{!}{
   \includegraphics{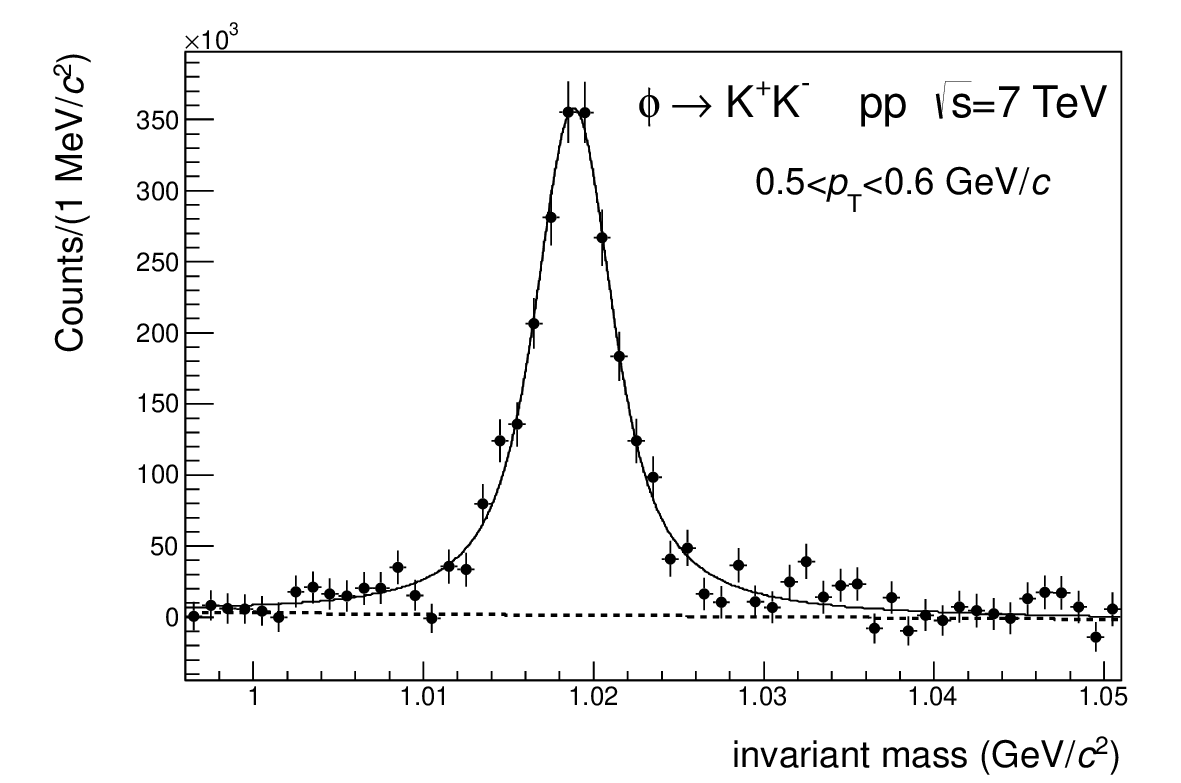}
}} 
\subfigure {\resizebox{0.5\textwidth}{!}{
   \includegraphics{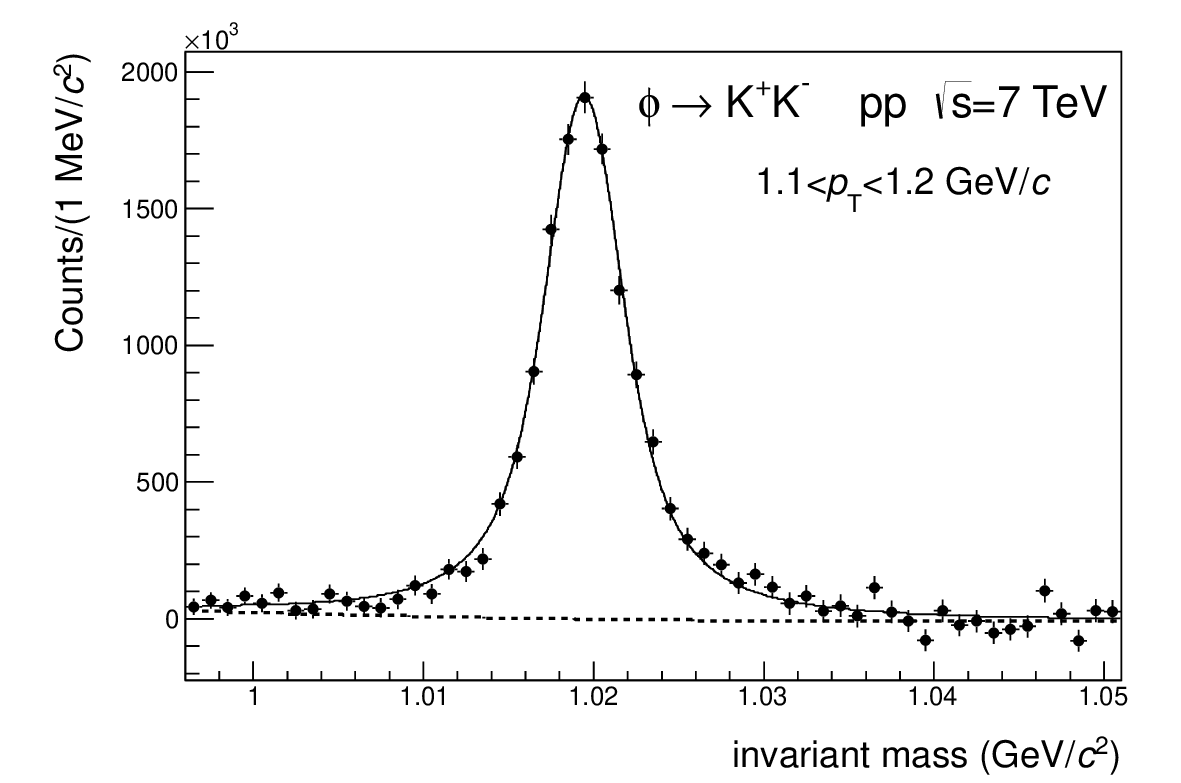}
}}
   
   \caption{(Colour online) (Upper panel) The  K$^+$K$^-$ invariant mass distribution  in 
   \modrap$<$0.5, for the bin 0.5$<$\pT$<$0.6~\gmom~(left) and 1.1$<$\pT$<$1.2~\gmom~(right)  in \pp~collisions at 7~TeV. 
   The solid curve is the fit result (Eq.~\ref{eqn:Voigtian}), 
   while the dashed line describes the background. The  background shape estimated using unlike-sign pairs from different events 
   (event mixing) or like-sign pairs from the same event are shown as open red squares and full green squares, respectively.   
   (Lower panel) The  K$^+$K$^-$ invariant mass distribution after mixed-event background subtraction for 0.5$<$\pT$<$0.6~\gmom~(left) 
   and 1.1$<$\pT$<$1.2~\gmom~(right). The solid curve is the fit result (Eq.~\ref{eqn:Voigtian}), 
   while the dashed line describes the residual background.
  }
   \label{fig:invmass_phi}

\end{figure*}

Considering the high multiplicities reached in \pp~collisions at LHC energies, good particle identification is important
to reduce combinatorial background as well as correlated background from misidentified resonance decays. 
The $\rmphi$ analysis requires only primary kaons to be selected and cuts were kept loose in order to maximize the efficiency.
The cut for particle identification in the TPC was set to 3$\sigma_{\mathrm {TPC}}$ (5$\sigma_{\mathrm {TPC}}$) for tracks with 
$p$ larger (smaller) than 0.35~\gmom.
When a TOF signal is present, a particle identification cut of $3\sigma_{\mathrm {TOF}}$ is also applied.
For the K$^*$  analysis, both pions and kaons are identified.   
Two different strategies were followed. 
For tracks with TOF signals, a TPC \dedx~cut of 5$\sigma_{\mathrm {TPC}}$ was applied and a TOF cut of 
3$\sigma_{\mathrm {TOF}}$ (2$\sigma_{\mathrm {TOF}}$) was applied for tracks with momenta below (above) 1.5~\gmom. 
For tracks without a TOF signal the kaon momentum was required to 
be below 0.7~\gmom and  5$\sigma_{\mathrm {TPC}}$,  3$\sigma_{\mathrm {TPC}}$, and  2$\sigma_{\mathrm {TPC}}$ cuts were used 
for $p<$0.35~\gmom, 0.35$<p<0.5$~\gmom, and $p>0.5$~\gmom, respectively.  This more restrictive cut on kaons 
was used to reduce the correlated background originating  from $\rm {\rho}$ decays in which a pion is misindentified.

\subsection{Raw yield extraction and background estimation}
\label{sec:raw}
The uncorrelated background was estimated using
two different techniques: like-sign and event mixing. In the
like-sign method invariant mass distributions of
 like-sign  K$\pi$ or KK combinations (for K$^*$ and $\rmphi$, respectively) from the same event were constructed.
In the event mixing method the shape of the uncorrelated background was estimated from the invariant  mass distribution of unlike-sign  
K$\pi$ or KK combinations from different events. To avoid  mismatch due to different acceptances and to assure a similar
event structure, tracks from events with similar vertex positions $z$ ($\Delta z <1$~cm) and track multiplicities $n$ ($\Delta n <10$) were mixed. To reduce statistical
uncertainties each event was mixed with 10 other events.
The mixed-event distribution was then normalized in the mass region 1.08 $<M<$1.2 (1.04 $<M<$1.07)~\gmass~for K$^*$ ($\rmphi$), and
subtracted in each {\pT} bin. 
The uncertainty in the normalization  was estimated by varying the normalization region and is included in the quoted systematic uncertainty for signal extraction. 
After background subtraction a residual background remains. This is due in part to  an imperfect
description of the combinatorial background but mainly caused by a real correlated background. The latter can arise from correlated $\pi$K or KK pairs or  
from misidentified particle decays  (for
example K$^{*0}$ for $\rmphi$, or $\rmphi$ and $\rho$ for K$^*$, or from underlying jet event structure).

The total \pT-integrated number of reconstructed  mesons after background subtraction was about 
1.8$\times$10$^6$ for~the~K$^*$ and 2.3$\times$10$^5$ for the $\rmphi$. 
For the K$^*$ the signal-to-background ratio varied from 0.08 at \pT=0.05~\gmom~to 0.2 at \pT= 5.5~\gmom.
The significance ($S/\sqrt{S+B}$) was about 34 in the \pT~bins at both 0.05 and 5.5~\gmom~and reached a maximum of about 127 at 
1~\gmom.
For the $\rmphi$ the signal-to-background ratio varied from 2.8 to 1.6 between \pT=0.45 and \pT=5.5~\gmom, with a minimum of 
0.5 at 1.6~\gmom; the significance was about 30 in the \pT~bins at both 0.45 and 5.5~\gmom~with a maximum of 90 at 1~\gmom. 

The raw yield of  \kstar~and its antiparticle was extracted in 22 \pT~bins between 0 and 6~\gmom~in the rapidity 
range \modrap$<0.5$. The combinatorial background was subtracted using like-sign $\pi^{\pm}$K$^{\pm}$ pairs.
The mass distribution $M$($\pi^{\pm}$,K$^{\mp}$)  (see Fig.~\ref{fig:invmass_Kstar} for two \pT~bins)  was fitted with 
a relativistic Breit-Wigner function multiplied by a Boltzmann 
factor~\cite{Adams:2005} and added to a  polynomial  residual background.  The width 
was found to be compatible, within uncertainties, with the natural value. 
At low \pT~the fitted mass values were found to be slightly lower (by about $\approx$5~\mmass) than the natural value, which 
is attributed to imperfections in corrections for the energy loss  in the detector material. 
To extract the yield the distribution of $M$($\pi^{\pm}$,K$^{\mp}$) 
 was then fitted 
with a (non-relativistic) Breit-Wigner function with the width fixed to its natural value ($\Gamma$= 48.7 $\pm$0.8~\cite{PDG2010}) and a background function:

\begin{eqnarray}
\frac{\mathrm{d}N}{\mathrm{d}M} = \frac{1}{2\pi}\frac{A\Gamma}{(M-M_0)^2+\Gamma^2/4}+ B(M)
   \label{eqn:Breit-Wigner}
\end{eqnarray}

\noindent where $A$  is the area under the peak corresponding 
to the number of K$^*$ mesons, $\Gamma$ is the full width at half maximum  of the peak, and $M_0$ is the resonance mass.
The residual background $B(M)$, after like-sign subtraction, was parametrized by a polynomial (dashed line in 
Fig.~\ref{fig:invmass_Kstar}).

\begin{figure}
\resizebox{0.8\textwidth}{!}{
   \includegraphics{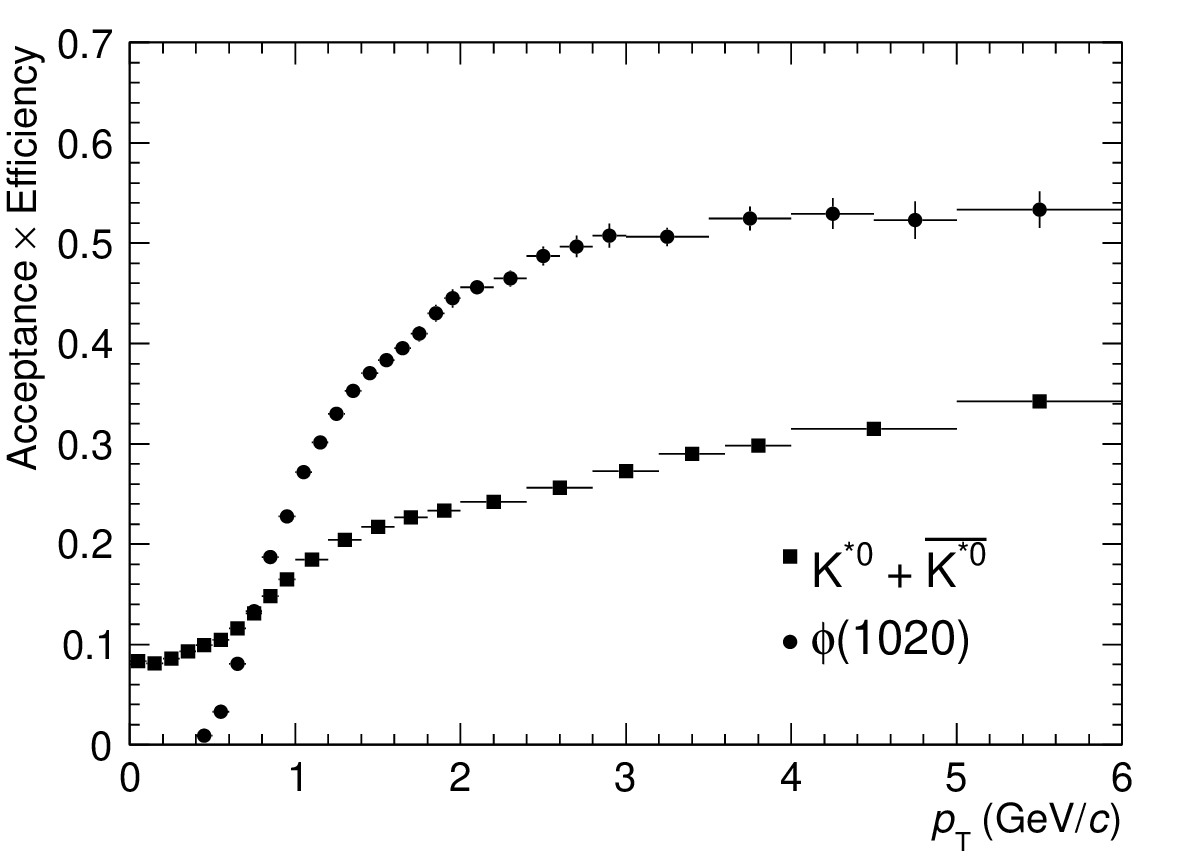}
}
\caption{ 
 The product of acceptance and efficiency of K$^*$ and $\rmphi$ detection as a function of \pT~in \modrap$<$0.5. Statistical uncertainties are reported. Contributions to the point-to-point systematic uncertainties are listed in Tab.~\ref{tab:1}.}
\label{fig:eff_phi_kstar}
\end{figure}
\begin{figure}
\resizebox{0.8\textwidth}{!}{
   \includegraphics{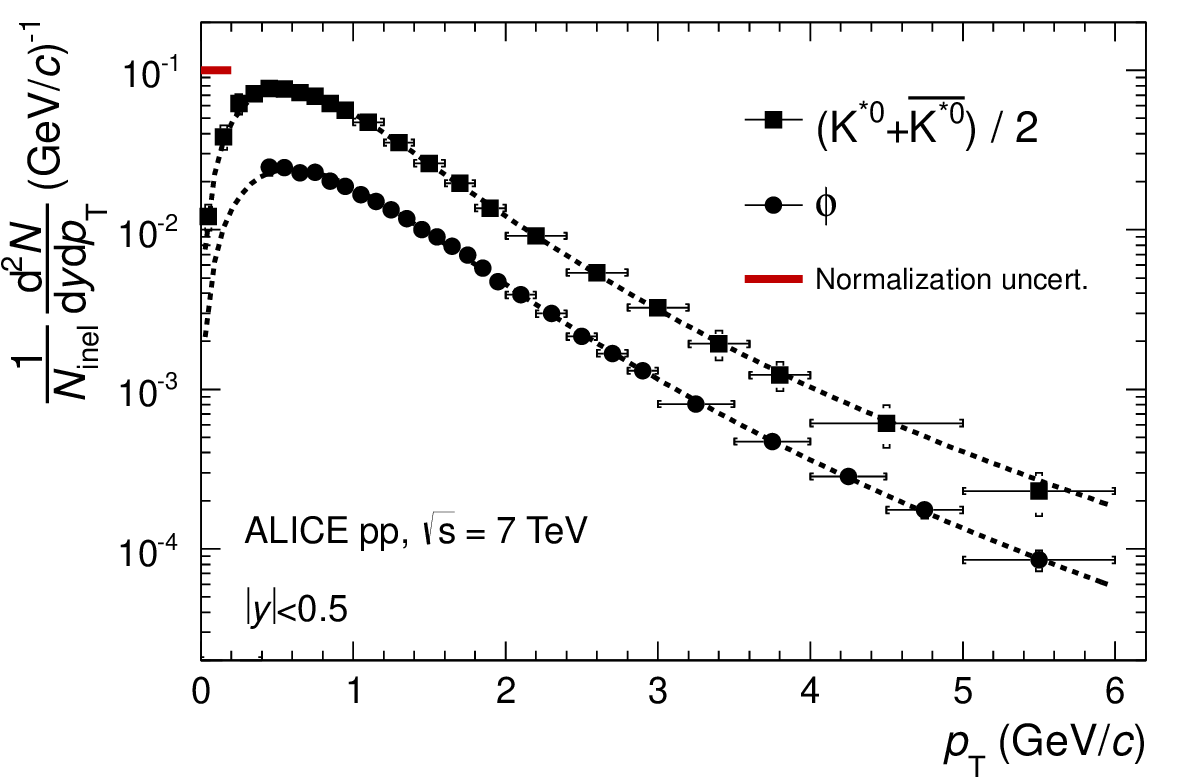}
}
\caption{Transverse momentum spectra for K$^*$ and \phir~in \pp~collisions at \s=7~TeV. 
The statistical and systematic uncertainties are added in quadrature 
and the uncertainty due to normalization~\cite{ALICE_cross_section}
is shown separately. 
The statistical uncertainty  is smaller than the symbol size. Each spectrum is fitted with a  L\'{e}vy-Tsallis function (dashed line).}
\label{fig:pt_spectra}
\end{figure}
For the $\rmphi$ meson, the raw yield was extracted from the K$^+$K$^-$ invariant mass distributions in 26 \pT~bins 
between 0.4 and 6~\gmom. The combinatorial background was subtracted using a polynomial fit (first or second order), like-sign pairs, 
or unlike-sign pairs from mixed events (Fig.~\ref{fig:invmass_phi} for two \pT~bins). 
Since the invariant mass resolution of the $\rmphi$ peak is of the same 
order of magnitude as the 
natural $\rmphi$ width ($\sim$1~{\mmass} vs. 4.26~\mmass), the fit is performed, after background subtraction, using 
a Voigtian function (convolution of Breit-Wigner function and Gaussian) superimposed on a polynomial to describe the residual
background: 

\begin{eqnarray}
\frac{\mathrm{d}N}{\mathrm{d}M} = A \int \frac{\Gamma/2 \pi}{(M-M^{\prime})^2+\Gamma^2/4} \frac{e^{-(M^{\prime}-M_0)^2/2\sigma^2}}{\sqrt{2\pi}\sigma} \mathrm{d}M^{\prime}  
+ B(M)
   \label{eqn:Voigtian}
\end{eqnarray}

\noindent where $\sigma$ represents the mass resolution and the other parameters have the same meaning as in 
Eq.~(\ref{eqn:Breit-Wigner}). 
The background $B$($M$) is represented in the lower panels  of Fig.~\ref{fig:invmass_phi} by a dashed line. 
The width $\Gamma$ is fixed to its nominal value~\cite{PDG2010} while $\sigma$ is a free parameter.
The fitted mass values were found to be compatible, within uncertainties, with the known mass~\cite{PDG2010}, 
with the exception of  the low \pT~range 0.4-0.7~\gmom~where a fitted value lower than the natural one  (by $<$ 0.1\%)  was observed.
The raw yields extracted using the three different methods to estimate the combinatorial background (analytic function, like-sign and mixed-event method) 
 were found to be compatible within a few percent; therefore the mean value of all three methods was taken in each \pT~bin.
\subsection{Efficiency corrections}
\label{sec:efficiency}
In order to extract the meson yields, the raw counts ($N^{\rm RAW}$) were corrected for 
the decay branching ratio~\cite{PDG2010} 
and for losses due to pion/kaon in-flight decays, geometrical acceptance, and detector efficiency 
($N^{\rm cor}=N^{\rm RAW}/(A\times\epsilon)$BR, where BR indicates the decay branching ratio).
The product of acceptance and efficiency ($A \times 
\epsilon$) was determined for K$^*$ and  $\rmphi$ from Monte Carlo 
simulations with the PYTHIA 6.4 event generator (tune Perugia~0~\cite{Perugia_tunes}) 
and  a GEANT3-based simulation of the ALICE detector response. 
About 60 M Monte Carlo events, with the same vertex distribution as the measured events,
were analyzed in exactly the same way as the data.
The dependence on the event generator was estimated to be below 1\% by comparing PYTHIA and PHOJET simulated events.
The $A\times\epsilon$  was determined from the Monte Carlo simulations as the ratio of the number of  reconstructed resonances to the number of those generated, 
differentially as a function of rapidity and transverse momentum. 
The transverse momentum dependence is shown in Fig.~\ref{fig:eff_phi_kstar}
 for  K$^*$ and  $\rmphi$ mesons.  
 The decrease in $A\times\epsilon$ at low \pT~is due to the minimum \pT~requirement for reconstructed tracks, while the different behaviour 
 for $\rmphi$  and  K$^*$ is due to the different $Q$-value of their decay (31.1 MeV for $\rmphi$ and 262.7~MeV for K$^*$).

Finally, corrections for the trigger efficiency ($\epsilon_{\rm trigger}$) and the required  primary vertex range ($\epsilon_{\rm vert}$) were applied 
in order to obtain the absolute resonance yields per inelastic collision: 

\begin{equation} 
\frac{\mathrm{d}^2N} {\mathrm{d}y\mathrm{d}\pT} = \frac{N^{\rm cor}(\pT)} 
{\Delta y \Delta \pT} \times \frac {1} {\epsilon_{\rm vert} }\times 
\frac{\epsilon_{\rm trigger}} {N_ {\rm MB}}
\end{equation}

\noindent here $N^{\mathrm {cor}}$ and  $N_ {\mathrm {MB}}$ are the number of reconstructed K$^*$ or $\rmphi$ and the total number of minimum bias triggers, respectively. The value of the  trigger selection efficiency for inelastic collisions
$\epsilon_{\mathrm {trigger}}$ is reported in Section~3. 
The loss of resonances due to the trigger selection, estimated by Monte Carlo, is negligible, less than 0.2\%.
The  $\epsilon_{\rm vert}$  correction factor accounts for resonance losses  ($\approx$1\%) due to the  requirement to have a vertex  in the range of $\pm$10~cm.

\begin{figure}
\resizebox{0.8\textwidth}{!}{    
   \includegraphics{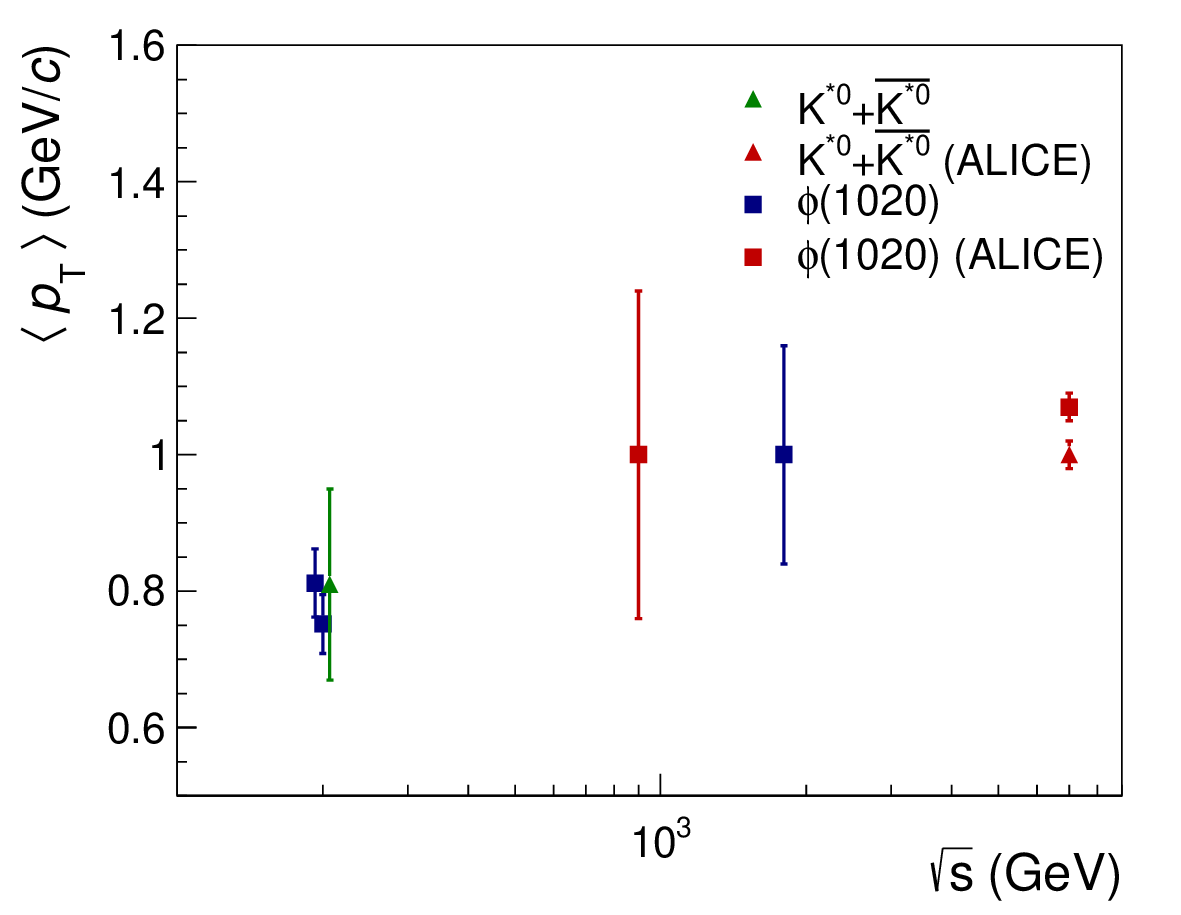}
}   
   \caption{ Energy dependence of \meanpT~for K$^*$ (triangles) and $\rmphi$ (squares) in \pp~collisions. 
   The points at lower energies are from STAR and PHENIX (\s=200~GeV)~\cite{Adams:2005,Abelev:2009,PHENIX_pp_phi}, ALICE (\s=0.9~TeV)~\cite{Strangeness_900}
   and E735 (\s=1.8~TeV)~\cite{Phi1800}. The STAR data have been slightly displaced to separate the K$^*$ and the $\rmphi$.
    The data point at 1.8~TeV represents the mean of the two values quoted from the E735 collaboration in~\cite{Phi1800}, obtained from two different fit functions of the $\rmphi$ 
\pT~distribution.}
   \label{fig:meanpt}
\end{figure}
\subsection{Estimation of the systematic uncertainties}
\label{sec:systematics}

The minimum and maximum values of the major contributions to the point-to-point systematic uncertainties are 
listed in Tab.~\ref{tab:1}.
The uncertainty due to the  raw yield extraction method was found to be  $\pm$2-28\% (2-10\%) for K$^*$ ($\rmphi$).
It was estimated   by changing the mass range considered for the fit and  the order of the polynomial for the residual 
 background function (from first through  third (second) order for  K$^*$ ($\rmphi$)). Finally, variations in the yield 
 due to the method used to estimate the combinatorial background (like-sign and event-mixing method and also analytic function for $\rmphi$) 
  were incorporated into the systematic uncertainties. 
For the K$^*$  a relativistic Breit-Wigner function was used to fit the mass peak in addition to the
non-relativistic version.  In the case of the  K$^*$ a rather large systematic uncertainty was estimated for the higher \pT~bins, due 
to the presence of a correlated background.

\begin{table}
\caption{Summary of the systematic point-to-point uncertainties in the K* and $\phi$  yield}
\label{tab:1}       
%
\begin{center}
\begin{tabular}{lcc}
\hline\noalign{\smallskip}
 Source of uncertainty & K$^*$  & $\phi$ \\ 
\noalign{\smallskip}\hline\noalign{\smallskip}
Signal extraction  &$\pm$ 2-28 \%& $\pm$ 2-10\% \\
Tracking efficiency &$\pm$ 8 \%& $\pm$ 8 \%\\
PID efficiency  &$\pm$  1-6 \% & $\pm$  1.5 \%\\
TOF matching efficiency &$\pm$  1-4 \% & - \\ 
\noalign{\smallskip}\hline\noalign{\smallskip} 
\end{tabular}
\end{center}
\end{table}

The uncertainty introduced by the  tracking and PID efficiency was estimated to be  $\pm$8\% (8\%) and  $\pm$1-6\% (1.5\%) 
respectively in the case of K$^*$ ($\rmphi$) by varying the kinematical 
and PID cuts on the daughter tracks. An additional  $\pm$1-4\% uncertainty was added for the K$^*$ due to 
differences observed in the TOF matching efficiency between data and Monte Carlo. 
The uncertainty on the yield contained in the extrapolated part of the $\rmphi$ spectrum was estimated to be 
$\pm$20\% using different fit functions.
The normalization to the number of inelastic collisions leads to a +6\% and -3\% uncertainty in the yield of the measured particles.
The resulting overall systematic uncertainty is $^{+10}_{-9}$\% ($^{+12.5}_{-11}$\%)
for the K$^*$ ($\rmphi$) yield
 \dndy~and $\pm$2\% (3\%) for the average transverse momentum \meanpT.
\begin{figure}
\resizebox{0.7\textwidth}{!}{
  \includegraphics{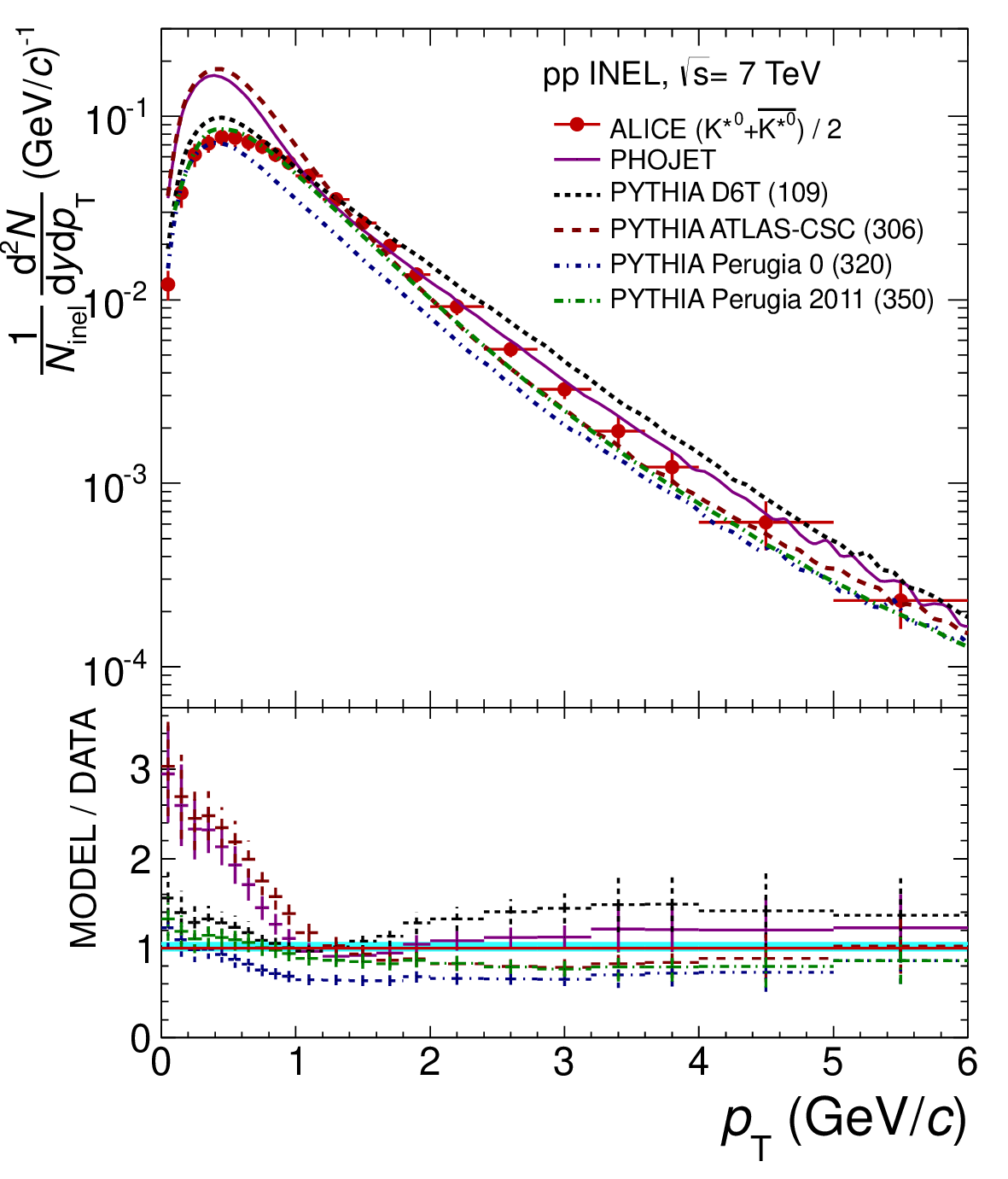}
}
\caption{(Colour online) Comparison of the K$^*$ \pT~spectra in inelastic \pp~collisions with PHOJET and  PYTHIA tunes D6T (109), 
ATLAS-CSC (306),  Perugia~0 (320), and Perugia~2011 (350).}
\label{fig:kstarmc}
\end{figure}
\begin{figure} 
\resizebox{0.7\textwidth}{!}{
 \includegraphics{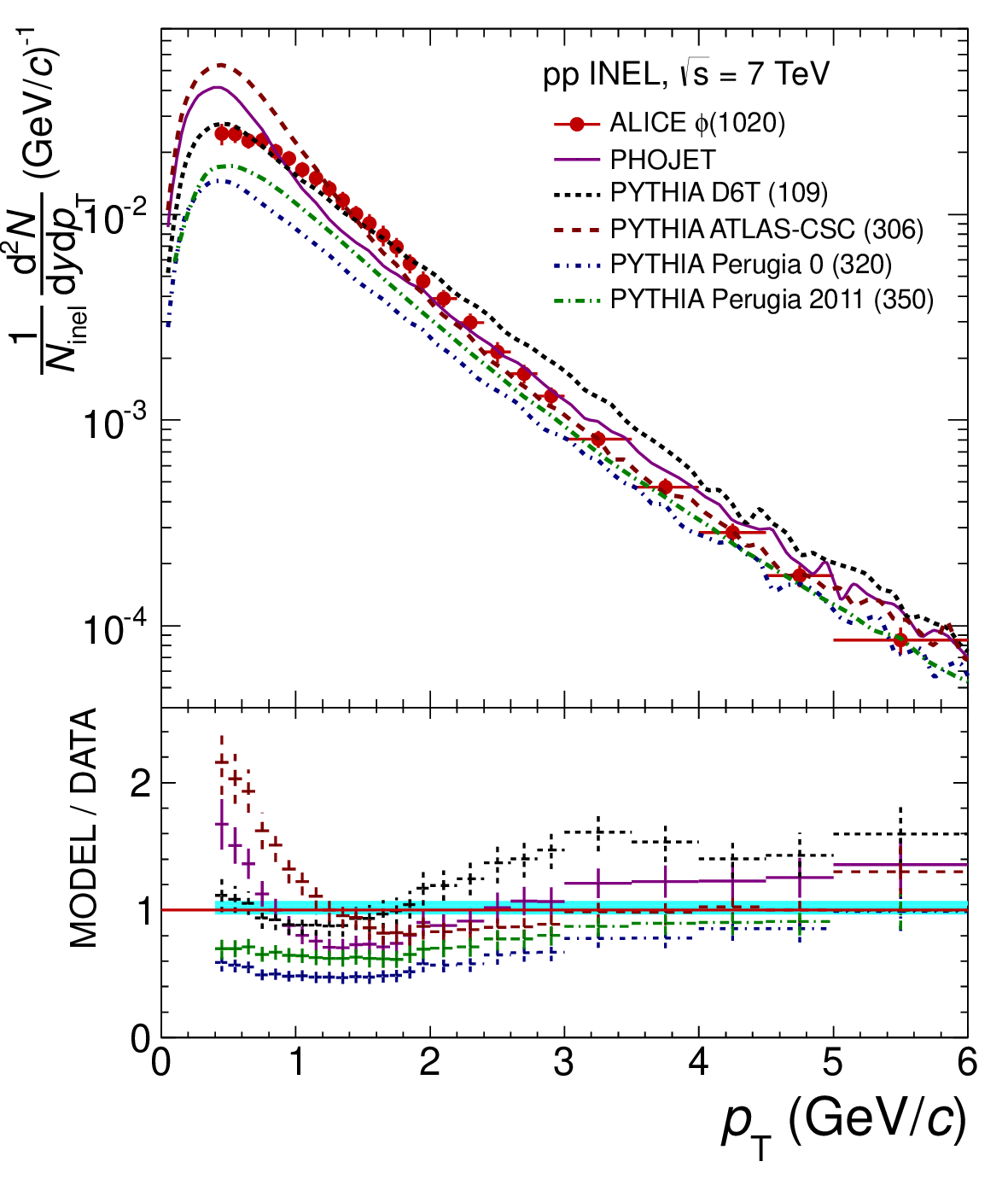}
}
\caption{(Colour online) Comparison of the \phir~\pT~spectra in inelastic \pp~collisions with PHOJET and PYTHIA tunes D6T (109), 
ATLAS-CSC (306), Perugia~0 (320), and Perugia~2011 (350).}
\label{fig:phimc}
\end{figure}
\section{Results and discussion}
\label{sec:inclusive}

\subsection{{ p$\mathbf {_T}$}~spectra and integrated yield }
\label{sec:ptspectra}

Figure~\ref{fig:pt_spectra} presents the corrected \pT~spectra for the two resonances. 
The statistical and point-to-point systematic uncertainties added in quadrature are shown. 
The spectra are fitted  with a L\'{e}vy-Tsallis function~\cite{Tsallis:1988,Abelev:2007}
\begin{equation}
%
\frac{\mathrm{d}^{2}N}{\mathrm{d}y\mathrm{d}\pT}   =  \frac{(n-1)(n-2)} {nT [nT + m(n-2)]} \times \frac{\mathrm{d}N}{\mathrm{d}y} \times {\pT} \times \left( 1+ \frac{m_{\mathrm{T}}-m}{nT}\right)^{-n}
\label{eqn:funclevy}
\end{equation}
where $m_{\rm T}=\sqrt{m^2+\pT^2}$.
This function describes both the exponential shape of the spectrum at 
low \pT~and the power-law distribution at large \pT, quantified by the inverse slope parameter $T$ and the 
exponent parameter $n$, respectively.
The extracted parameter values are listed in Table~\ref{tab:2} and the fits are shown in Fig.~\ref{fig:pt_spectra}. The $\chi^2$/ndf values are smaller than unity because the point-to-point systematic uncertainties, which are included in the fit, could be correlated. 

The extracted $n$ values 
are similar to those quoted by the STAR experiment at RHIC for the
$\rmphi$ measured in \pp~collisions at 200~GeV ($n$=8.3$\pm$1.2)~\cite{Abelev:2009}. 
In contrast, the slope parameters are significantly higher than the values obtained at RHIC,
$T$= 202 $\pm$ 14 $\pm$ 11~MeV for $\rmphi$, and $T$=223 $\pm$8$\pm$9~MeV for K$^*$~\cite{Adams:2005} 
(the latter was obtained by an exponential fit and can therefore not be directly compared).

The total yields \dndy~and the mean transverse momentum \meanpT, including statistical and systematic uncertainties, are listed 
in Table~\ref{tab:3}.
The values of \dndy~were obtained by integrating the spectra in the measured range and extrapolating 
to zero \pT~with the fitted L\'{e}vy-Tsallis function.
The contribution of the low-\pT~extrapolation is negligible for the K$^*$ and 
about 15 $\pm$ 3\% for the $\rmphi$.
 The mean transverse momentum was estimated in the range 0$<\pT<6$~\gmom~using  the L\'{e}vy-Tsallis function. 
However, similar values are obtained when calculating the mean from the measured data points, using the fit only to extrapolate
into the unmeasured \pT~regions. In addition to  
the point to point systematic uncertainties previously described, an exponential fit was also used to estimate the  systematic uncertainty in {\meanpT} 
due to a different choice of fit function.
Compared to pp collisions at 200 GeV~\cite{Adams:2005,Abelev:2009,PHENIX_pp_phi}, the mean \pT~rises by about 30\% (Fig.~\ref{fig:meanpt}) and 
the yield per inelastic collision increases by about a factor of two, which is similar to the overall increase of charged particle multiplicity~\cite{ALICE_multiplicity_23,ALICE_multiplicity7}.
 
\begin{table*}
\caption{Parameters extracted from the L\'{e}vy-Tsallis~(\ref{eqn:funclevy}) fits to the $\rm{K}^{*}$ and $\phi$~transverse 
momentum spectra in 7 \tev~\pp~collisions,  including 
point-to-point systematic uncertainties. The first uncertainty  is statistical and the
second is systematic.}
\label{tab:2}       
\begin{center}
\begin{tabular}{cccc}
\hline\noalign{\smallskip}
{Particles} 	       &  $\chi^2$/ndf           &  $T$~(MeV)             & $n$	\\
\noalign{\smallskip}\hline\noalign{\smallskip}
  K$^*$    & $2.0/19$    & $254 \pm 2 \pm 18$  & $ 6.2 \pm 0.07 \pm 0.8$    \\ 
 $\phi$     & $2.8/23$   & $272 \pm 4 \pm  11$  & $ 6.7 \pm 0.20 \pm 0.4$   \\ 
					                                        
\noalign{\smallskip}\hline 
\end{tabular}
\end{center}
\end{table*}

%
\begin{table*}
\caption{$\rm{K}^{*}$ and $\rm{\phi}$ yield and {\meanpT} estimated in the range 0-6 \gmom~in inelastic pp collisions at \s=7~TeV.
The systematic uncertainties of \dndy~and \meanpT~include  contributions  from the choice of spectrum fit function for extrapolation, the absolute normalization,
and the point-to-point uncertainties listed in Tab~\ref{tab:1}.}
\label{tab:3}       
\begin{center}
\begin{tabular}{ccccccc}
\hline\noalign{\smallskip}
{Particles}                   & {measured \pT}  (\gmom) &
\dndy                        & {\meanpT} (\gmom)         \\
\noalign{\smallskip}\hline\noalign{\smallskip}
 K$^*$  & $[0.0-6.0]$  & $0.097  \pm 0.0004  ^{+0.010} _{-0.009}$    & $1.01  \pm 0.003
 \pm 0.02$  \\
 $\phi$     & $[0.4-6.0]$      & $0.032 \pm 0.0004 ^{+0.004} _{-0.0035}$ & $1.07 \pm 0.005 \pm 0.03$   \\
\noalign{\smallskip}\hline
\end{tabular}
\end{center}
\end{table*}

The $\rmphi$ yield, measured via the leptonic decay channel in the ALICE muon spectrometer in 2.5$<y<$4,  
1$<$\pT$<$5~\gmom~\cite{Phimumu}, has a similar momentum distribution, but 
is lower by about 30\% at  forward rapidity.
The  $\rmphi$ yield is expected to vary by  20\%-50\% between forward (2.5$<y<$4) and mid-central (-0.5$<y<$0.5) rapidities, based
on analysis of different PYTHIA tunes described in  paragraph~\ref{sec:models}.
 In particular, the lower value is predicted from the D6T PYTHIA tune~\cite{D6T}, which reproduces rather well the $\rmphi$ spectrum at forward rapidity~\cite{Phimumu} and the low \pT~part of the $\rmphi$ spectrum at mid-rapidity (see Fig.~\ref{fig:phimc} described in~\ref{sec:models}).  

\begin{figure}
\resizebox{0.7\textwidth}{!}{
 \includegraphics{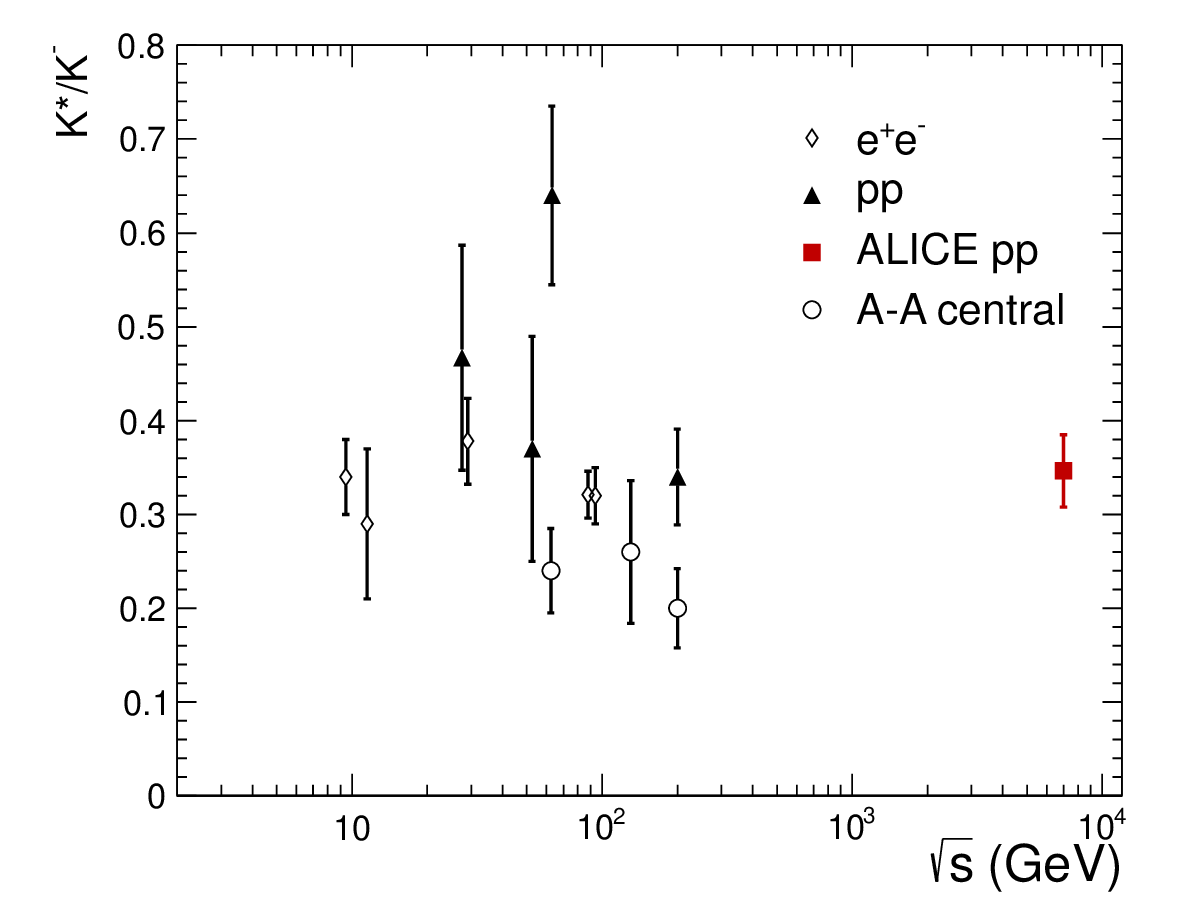}
} 

\resizebox{0.7\textwidth}{!}{
   \includegraphics{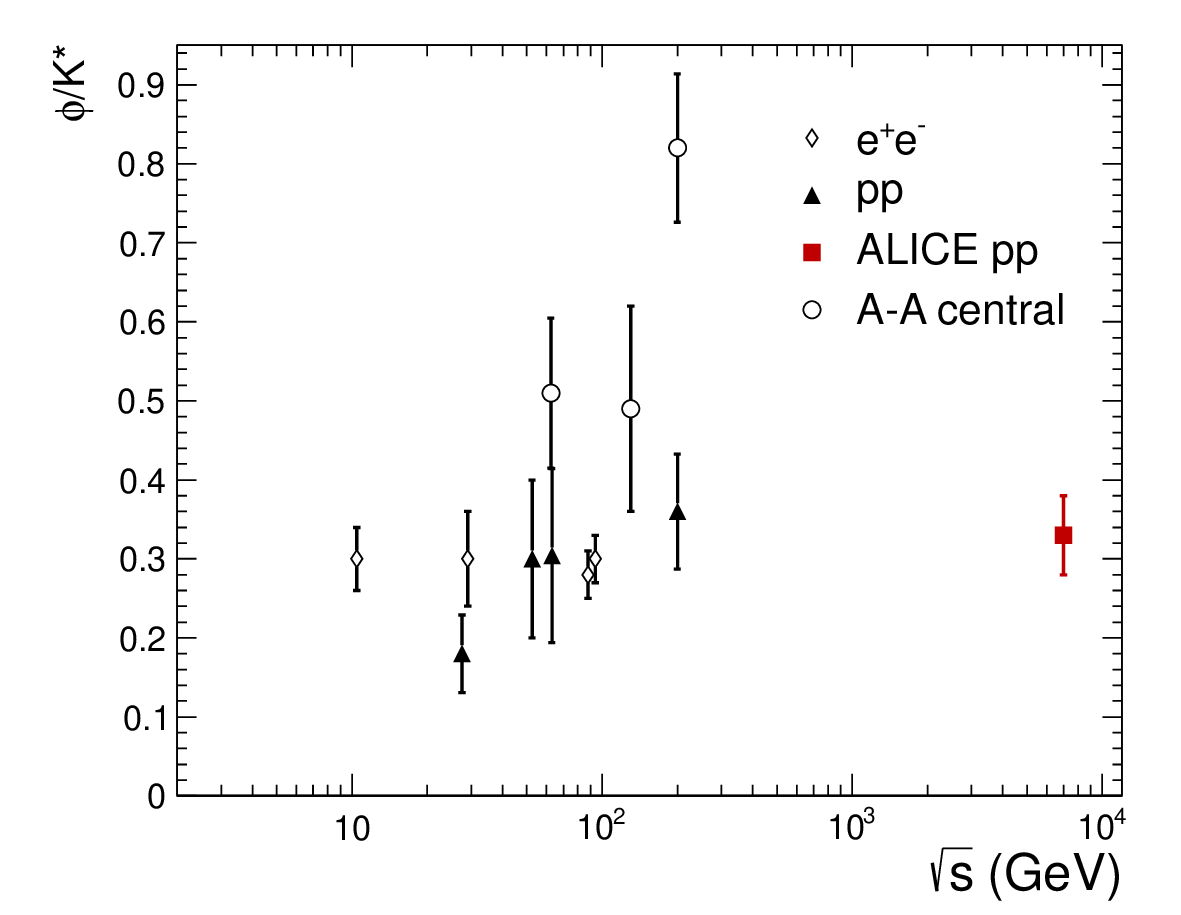}
}   
   \caption{(Colour online) Energy dependence of the  K$^*$/K$^-$ (upper panel) and $\rmphi$/K$^*$~(lower panel) ratio in 
   e$^+$e$^-$ (diamonds)~\cite{Albrecht:1994,Pei:1996,Hofmann:1988,Abe:1999,Behrends:1985}, 
   and \pp~(triangles)~\cite{Aguillar:1991,Adams:2005,Abelev:2009,Drijard:1981,Akesson:1982,Aggarwal:2011} collisions. 
   Red squares represent the data from the ALICE experiment for 7~\tev~\pp~collisions, 
   K$^-$ yields are from~\cite{Spectra_7}. 
   Open circles represent the same ratios in central nucleus-nucleus collisions 
   from~\cite{Adams:2005,Abelev:2009,Aggarwal:2011,Adler:2002a,Adler:2002b}.
   Some points have been displaced horizontally for better visibility.
   Ratios are calculated from yields at mid-rapidity or in full phase space.
 }
   \label{fig:comparisons1}
\end{figure}

\begin{figure}
\resizebox{0.7\textwidth}{!}{    
  \includegraphics{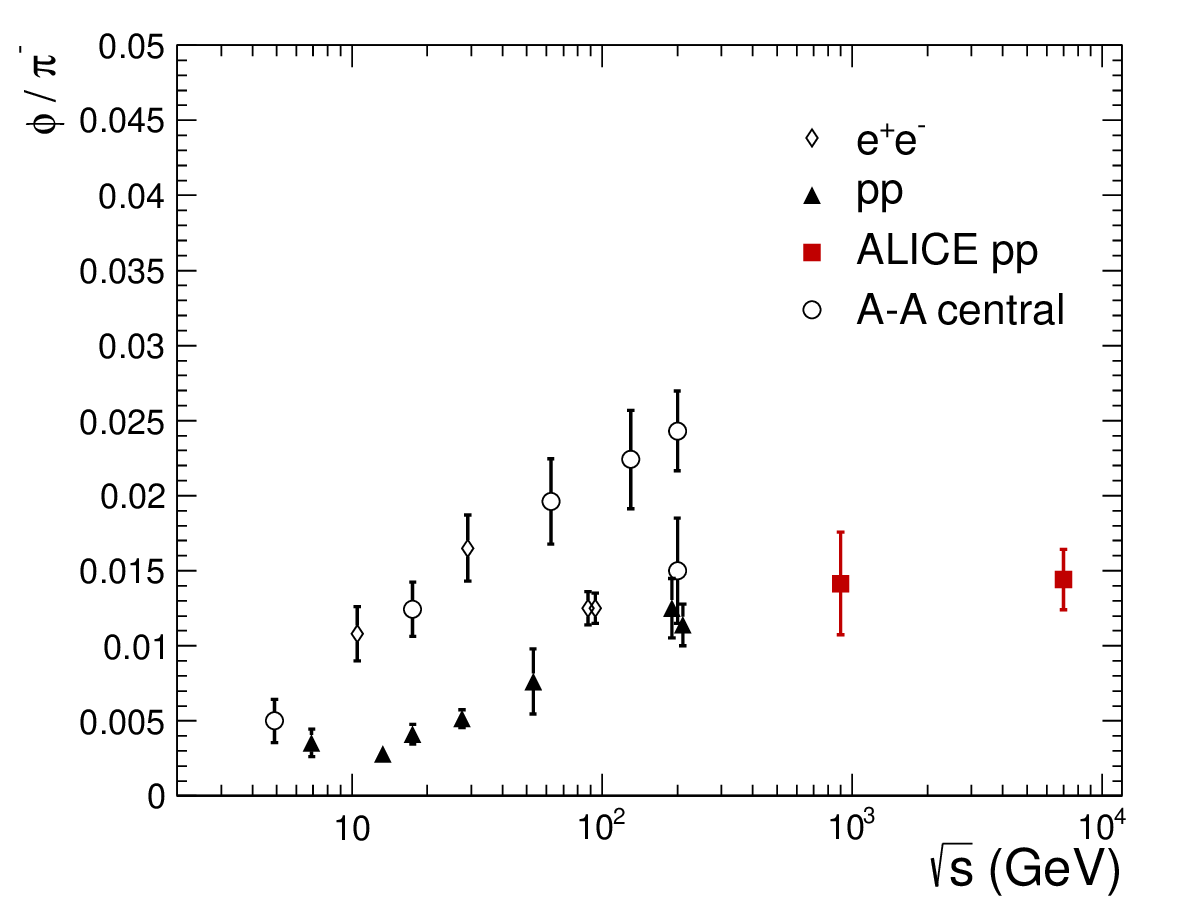}   
}
\resizebox{0.7\textwidth}{!}{
  \includegraphics{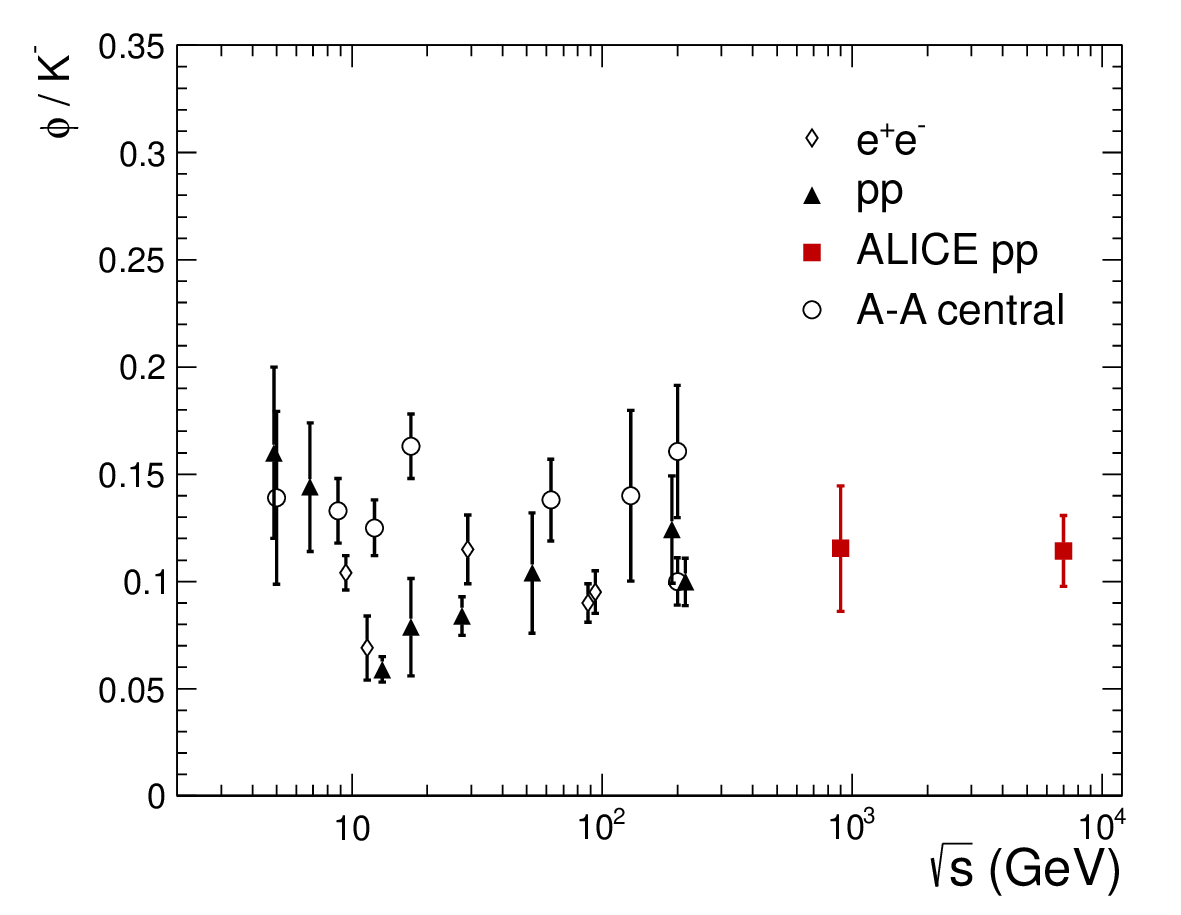}
}    
   \caption{(Colour online) Energy dependence of the $\rmphi/\pi^{-}$ (upper panel) and $\rmphi/$K$^-$ ratio (bottom panel) in nuclear (open circles)~\cite{Adler:2005,Abelev:2009,Adler:2002a,Back:2004,Afanasiev:2005,Alt:2008,Adcox:2002}, e$^+$e$^-$ (diamonds)~\cite{Albrecht:1994,Pei:1996,Hofmann:1988,Abe:1999,Behrends:1985}, 
   and {\pp} (triangles)~\cite{Aguillar:1991,Adams:2005,Abelev:2009,PHENIX_pp_phi,Blobel:1975,Drijard:1981,Back:2004,Afanasiev:2005,Alt:2008,Daum:1981}
   collisions.  Other $\rm \pi^{-}$ and K$^-$ yields are from ~\cite{Adams:2004,Daum:1981,Rossi:1975,PHENIX_pp}.
    Red squares represent the ALICE data at 0.9 and 7~TeV. The pion and kaon yields at 7~TeV are 
    from~\cite{Spectra_7}. The $\rmphi$, $\pi^-$, and K$^-$ yields at 0.9~TeV are from~\cite{Spectra_900,Strangeness_900}.
    Some points have been displaced horizontally for better visibility.
   Ratios are calculated from yields at mid-rapidity or in full phase space, except the data at \s=4.87~ GeV~\cite{Back:2004}. 
    }
   \label{fig:comparisons2}
\end{figure}

\subsection{Comparisons to models}
\label{sec:models}

Multiparticle production, which is predominantly a soft, non-pertubative process, is usually modelled by QCD inspired 
Monte Carlo event generators like PHOJET~\cite{PHOJET} and PYTHIA~\cite{PYTHIA}. In both models, hadronization is simulated using 
the Lund string fragmentation model~\cite{LundModel}.
Different PYTHIA tunes were obtained by adjusting the model  parameters to reproduce existing data. 
The D6T tune~\cite{D6T}, which uses the CTEQ6L parton distribution function (with a corresponding larger production of 
strange particles), was obtained by fitting CDF Run 2 data. The ATLAS-CSC~\cite{ATLAS} tune was adjusted to minimum bias 
data from the UA5, E735, and CDF experiments for energies ranging from 0.2 to 1.8 TeV.
The latest PYTHIA tune,  Perugia~2011~\cite{Perugia_tunes}, takes into account first results  from  the LHC, in particular 
minimum-bias and underlying event data  at 0.9 and 7~TeV. Strange baryon production 
was increased in this tune leading to a larger \rmLambda/K ratio with respect to the Perugia~0 tune.

The transverse momentum spectra of K$^{*}$ and $\rmphi$ are compared to PHOJET and various PYTHIA tunes in 
Figs.~\ref{fig:kstarmc} and~\ref{fig:phimc}. 
For PYTHIA, tunes D6T (109), ATLAS-CSC (306), Perugia~0 (320) and  Perugia~2011 (350) were used.
The best agreement is found for the PYTHIA Perugia~2011 tune, which reproduces both the K$^{*}$ spectrum and the 
high \pT~part (\pT $>$ 3~\gmom) of the $\rmphi$ spectrum rather well. PHOJET and ATLAS-CSC very significantly 
overestimate the low momentum part (\pT $<$ 1~\gmom) of the transverse momentum distribution but 
reproduce the high momentum distribution of both mesons well. 
The PYTHIA D6T tune 
gives the best description at low \pT, but deviates from the data at \pT$>$2~\gmom.
Finally, the PYTHIA Perugia~0 tune underestimates the meson yield for \pT~larger than 0.5~\gmom. 

Similar comparisons for the mid- and forward-rapidity $\rmphi$ spectrum in pp collisions at \s=0.9~TeV~\cite{Strangeness_900} and 7~TeV~\cite{Phimumu}, respectively, 
show that the  $\rmphi$ spectrum is rather well reproduced 
by the ATLAS-CSC and D6T tunes, while the Perugia 0 and 2011 tunes underestimate the data. Moreover the PYTHIA tunes generally underestimate strange meson 
and hyperon production in 7~TeV pp collisions ~\cite{Omega_ALICE,CMS_7}, while the Perugia 2011 tune gives a good description of kaon production in pp collisions at 7~TeV~\cite{Spectra_7}.

\subsection{Particle ratios}
\label{sec:comparison}

The measurement of particle production and particle ratios in \pp~collisions is important as a baseline for comparison with heavy ion 
reactions.
In heavy ion collisions, the yields for stable and long-lived hadrons reflect the thermodynamic conditions 
(temperature, chemical potentials) at freeze-out, whereas the yield for short-lived resonances 
can be modified by final state interactions inside the hot and dense reaction zone~\cite{Markert2005,Bleicher2002}. 
 Particularly interesting is the comparison of $\rmphi$ and K$^*$ production, considering the different lifetimes 
(about a factor 10) of the two resonances.

Using different particle ratios
(like K/$\pi$ or $\rmphi$/K$^{*}$) measured in elementary collisions, values ranging from 0.1 
to 0.4~\cite{Aguillar:1991,Albrecht:1994,Pei:1996,Hofmann:1988,UA1_strange} were previously obtained for the
strange quark suppression factor $\lambda_{\mathrm {s}}= 2\mathrm{s}\bar{\mathrm{s}}/(\mathrm{u}\bar{\mathrm{u}}+\mathrm{d}\bar{\mathrm{d}})$, which represents
the probability to produce strange quark pairs relative to light quarks~\cite{Becattini:1998}.   
In \pp~reactions, particle abundances have been successfully described by statistical-ther-
\noindent mal models. 
Now, using measured identified particle yields, an energy-independent value of 0.2 for  
$\lambda_{\mathrm {s}}$ 
has been 
extracted in e$^+$e$^-$, \pp, and p$\overline{\mathrm{p}}$ collisions at \s$<$1~TeV~\cite{Becattini:1998,Becattini:2001}.

\begin{figure}
\resizebox{0.8\textwidth}{!}{   
   \includegraphics{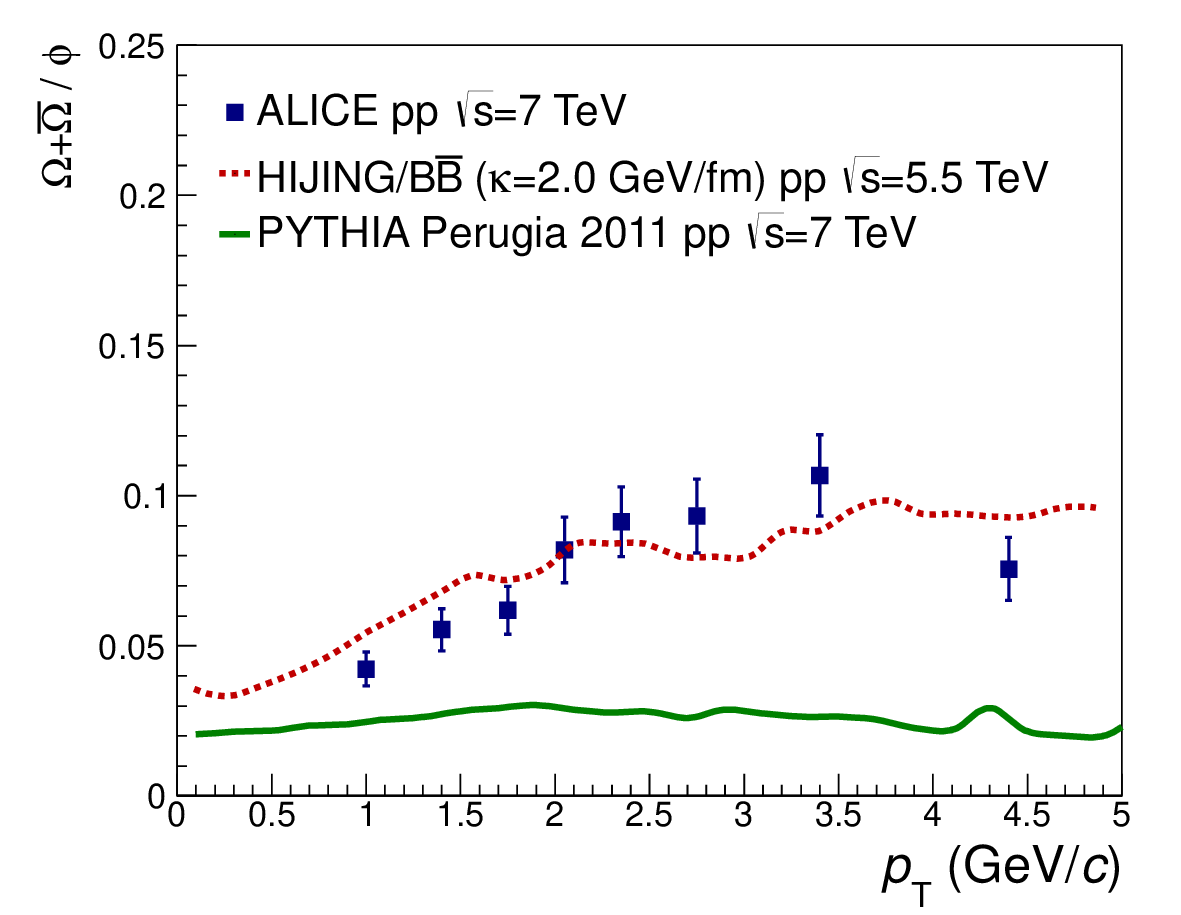}
}   
   \caption{ (Colour online) $(\rmOmega^+ + \rmOmega^{-})/\rmphi$ ratio as a function of transverse momentum for 
   \pp~collisions at \s=7~TeV.
   $\rmOmega$ data are from~\cite{Omega_ALICE}. The dashed line represents the prediction of HIJING/B$\overline{\rm B}$ v2.0 model with 
   a SCF for \pp~collisions 
   at \s=5.5~TeV with a string tension of 2~GeV/fm~\cite{LHCprediction}.
   The same calculation at 7~TeV yields a $\sim$ 10\% higher ratio~\cite{Topor_private}.
   The full line represents the prediction of the PYTHIA Perugia~2011 tune~\cite{Perugia_tunes} for 
   \pp~collisions at \s=7~TeV.}
   \label{fig:omega_phi_ratio_smooth}
\end{figure}

Using the $\rmphi$ and K$^*$ yields 
presented in this paper and stable particle results measured by ALICE
at the same energy~\cite{Spectra_7}, we find the following values for particle ratios in \pp~collisions at 7~TeV:
K$^*$/K$^-$= 0.35$\pm$ 0.001 (stat.) $\pm$0.04 (syst.),
$\rmphi$/K$^{*}$= 0.33 $\pm$0.004 (stat.) $\pm$0.05 (syst.),
$\rmphi/$K$^-$= 0.11 $\pm$0.001(stat.) $\pm$0.02 (syst.), 
$\rmphi/\pi^-$= 0.014 $\pm$0.0002 (stat.) $\pm$0.002 (syst.).
Due to the fact that the same data were analyzed to extract both resonance and non-resonance  ($\rm{\pi}$,K) yields,
 the uncertainties due to the absolute normalization cancel and are therefore not included in the systematic uncertainties of the ratios.
These ratios are shown in Figures~\ref{fig:comparisons1} and~\ref{fig:comparisons2},  together with the results obtained at 
lower incident energies in \pp, e$^+$e$^-$, and A-A collisions.

The K$^*$/K$^-$, $\rmphi/$K$^-$, and $\rmphi$/K$^{*}$ ratios are essentially independent of energy and also independent 
of the collision system, with the exception of K$^*$/K and $\rmphi/$K$^*$ at RHIC~\cite{Abelev:2009,Aggarwal:2011,Adler:2002a,Adler:2002b}, where these ratios in nuclear collisions 
are respectively lower and higher than in \pp.
On the contrary, the $\rmphi/\pi$ ratio increases with energy both in heavy ion and in \pp~collisions up to at least 
200 GeV. However,  in heavy ion collisions the  value  obtained by the PHENIX experiment~\cite{Adler:2005}, about  40\% lower than the STAR result~\cite{Abelev:2009} at the same  collision energy, 
seems indicate  a saturation of this ratio at the RHIC energies.
In \pp~collisions we observe a saturation of the  $\rmphi/\pi$ ratio, with
no significant change over the LHC energy range between 1 and 7~TeV.

In microscopic models where soft particle production is governed by string fragmentation, strange hadron yields are 
predicted to depend on the string tension~\cite{Toporpop:2011}. 
Multi-strange baryons, and in particular the ratio $\rmOmega/\rmphi$, are expected to be very sensitive to this 
effect~\cite{LHCprediction}. 
The $\rmphi$ yield is compared to the $\rmOmega^- + \overline{\rmOmega}^+$ data measured by ALICE at the same incident 
energy~\cite{Omega_ALICE} in Fig.~\ref{fig:omega_phi_ratio_smooth} as a function of transverse momentum. 
The full line represents the PYTHIA model (Perugia~2011 tune),  which is a factor 1.5-5 below the data. 
While this tune describes the $\rmphi$ spectrum reasonably well above 2-3 \gmom, it underpredicts multistrange baryon 
yields by a large factor~\cite{Omega_ALICE}. 
The dashed line, which is very close to the data,  represents the prediction of a model with increased string tension, 
the HIJING/B$\overline{\rm B}$ v2.0 model with a Strong Colour Field (SCF), for \pp~collisions at 5.5~\tev~\cite{LHCprediction}. This is a model that 
combines multiple minijet production via perturbative QCD  with soft longitudinal string excitation and hadronization.  In this case the SCF effects are modeled by varying the effective string tensions that controls the \qqbar~and \qqqqbar~pair creation rates and the strangeness suppression factor. 
The value of string tension used in this calculation is $\kappa$=2 GeV/fm, equal to the value used to fit 
the high baryon/meson ratio at \s=1.8~\tev~reported by the CDF collaboration~\cite{CDF151}.
The same calculation at 7~TeV yields a $\sim$ 10\% higher ratio~\cite{Topor_private}.  
Higher values of the string tension  ($\sim$ 3 GeV/fm) also successfully  reproduce also the
($\rmOmega + \overline{\rmOmega})/\rmphi$ ratio in  Au-Au collisions at \s=200~GeV~\cite{LHCprediction}, but overestimate the (\rmLambda + \rmAlambda)/\Kzs~at 
7~TeV~\cite{Toporpop:2011}.

\section{Conclusion}
\label{sec:conclusion}

Yields and spectra of {\kstar} and {\phir} mesons were measured for inelastic \pp~collisions 
at \s=7~TeV by the ALICE collaboration at the LHC. The transverse momentum spectra are well described 
by the L\'{e}vy-Tsallis function. The yields for both mesons increase by about a factor of two from 
200 GeV centre-of-mass energy, and the average \pT~by about 30\%. 
 
The K$^*$/K and $\rmphi$/K$^*$ ratios (and consequently the $\rmphi$/K ratio) are found to be independent of energy up to 
7~TeV. Also the $\rmphi/\pi$ ratio, which increases in both \pp~and A-A collisions up to at least RHIC energies,  
saturates and becomes independent of energy above 200 GeV.

The data have been compared to a number of PYTHIA tunes and the PHOJET event generator. None of them gives a fully 
satisfactory description of the data. The latest PYTHIA version (Perugia~2011) comes closest, while  still 
underpredicting the $\rmphi$ meson \pT~spectrum below 3 \gmom~by up to a factor of two.

The $(\rmOmega^- + \overline{\rmOmega}^+)/\rmphi$ ratio is not  reproduced by PYTHIA Perugia~2011, but is in good agreement 
with the HIJING/B$\overline{\rm B}$ v2.0 model with SCF, which enhances multi-strange baryon production 
by increasing the string tension parameter.

\newenvironment{acknowledgement}{\relax}{\relax}
\begin{acknowledgement}
\section{Acknowledgements}
The ALICE collaboration would like to thank all its engineers and technicians for their invaluable contributions to the construction of the experiment and the CERN accelerator teams for the outstanding performance of the LHC complex.
\\
The ALICE collaboration acknowledges the following funding agencies for their support in building and
running the ALICE detector:
 \\
Calouste Gulbenkian Foundation from Lisbon and Swiss Fonds Kidagan, Armenia;
 \\
Conselho Nacional de Desenvolvimento Cient\'{\i}fico e Tecnol\'{o}gico (CNPq), Financiadora de Estudos e Projetos (FINEP),
Funda\c{c}\~{a}o de Amparo \`{a} Pesquisa do Estado de S\~{a}o Paulo (FAPESP);
 \\
National Natural Science Foundation of China (NSFC), the Chinese Ministry of Education (CMOE)
and the Ministry of Science and Technology of China (MSTC);
 \\
Ministry of Education and Youth of the Czech Republic;
 \\
Danish Natural Science Research Council, the Carlsberg Foundation and the Danish National Research Foundation;
 \\
The European Research Council under the European Community's Seventh Framework Programme;
 \\
Helsinki Institute of Physics and the Academy of Finland;
 \\
French CNRS-IN2P3, the `Region Pays de Loire', `Region Alsace', `Region Auvergne' and CEA, France;
 \\
German BMBF and the Helmholtz Association;
\\
General Secretariat for Research and Technology, Ministry of
Development, Greece;
\\
Hungarian OTKA and National Office for Research and Technology (NKTH);
 \\
Department of Atomic Energy and Department of Science and Technology of the Government of India;
 \\
Istituto Nazionale di Fisica Nucleare (INFN) of Italy;
 \\
MEXT Grant-in-Aid for Specially Promoted Research, Ja\-pan;
 \\
Joint Institute for Nuclear Research, Dubna;
 \\
National Research Foundation of Korea (NRF);
 \\
CONACYT, DGAPA, M\'{e}xico, ALFA-EC and the HELEN Program (High-Energy physics Latin-American--European Network);
 \\
Stichting voor Fundamenteel Onderzoek der Materie (FOM) and the Nederlandse Organisatie voor Wetenschappelijk Onderzoek (NWO), Netherlands;
 \\
Research Council of Norway (NFR);
 \\
Polish Ministry of Science and Higher Education;
 \\
National Authority for Scientific Research - NASR (Autoritatea Na\c{t}ional\u{a} pentru Cercetare \c{S}tiin\c{t}ific\u{a} - ANCS);
 \\
Federal Agency of Science of the Ministry of Education and Science of Russian Federation, International Science and
Technology Center, Russian Academy of Sciences, Russian Federal Agency of Atomic Energy, Russian Federal Agency for Science and Innovations and CERN-INTAS;
 \\
Ministry of Education of Slovakia;
 \\
Department of Science and Technology, South Africa;
 \\
CIEMAT, EELA, Ministerio de Educaci\'{o}n y Ciencia of Spain, Xunta de Galicia (Conseller\'{\i}a de Educaci\'{o}n),
CEA\-DEN, Cubaenerg\'{\i}a, Cuba, and IAEA (International Atomic Energy Agency);
 \\
Swedish Research Council (VR) and Knut $\&$ Alice Wallenberg
Foundation (KAW);
 \\
Ukraine Ministry of Education and Science;
 \\
United Kingdom Science and Technology Facilities Council (STFC);
 \\
The United States Department of Energy, the United States National
Science Foundation, the State of Texas, and the State of Ohio.
\end{acknowledgement}
\newpage
%
%
\appendix
\section{The ALICE Collaboration}
\label{app:collab}
\begingroup
\small
\begin{flushleft}
B.~Abelev\Irefn{org1234}\And
J.~Adam\Irefn{org1274}\And
D.~Adamov\'{a}\Irefn{org1283}\And
A.M.~Adare\Irefn{org1260}\And
M.M.~Aggarwal\Irefn{org1157}\And
G.~Aglieri~Rinella\Irefn{org1192}\And
A.G.~Agocs\Irefn{org1143}\And
A.~Agostinelli\Irefn{org1132}\And
S.~Aguilar~Salazar\Irefn{org1247}\And
Z.~Ahammed\Irefn{org1225}
% ...
\renewcommand\labelenumi{\textsuperscript{\theenumi}~}
\section*{Affiliation notes}
\renewcommand\theenumi{\roman{enumi}}
\begin{Authlist}
\item \Adef{M.V.Lomonosov Moscow State University, D.V.Skobeltsyn Institute of Nuclear Physics, Moscow, Russia}Also at: M.V.Lomonosov Moscow State University, D.V.Skobeltsyn Institute of Nuclear Physics, Moscow, Russia
\item \Adef{University of Belgrade, Faculty of Physics and "Vin{c}a" Institute of Nuclear Sciences, Belgrade, Serbia}Also at: University of Belgrade, Faculty of Physics and "Vin\v{c}a" Institute of Nuclear Sciences, Belgrade, Serbia
\end{Authlist}
\section*{Collaboration Institutes}
\renewcommand\theenumi{\arabic{enumi}~}
\begin{Authlist}
\item \Idef{org1279}Benem\'{e}rita Universidad Aut\'{o}noma de Puebla, Puebla, Mexico
\item \Idef{org1220}Bogolyubov Institute for Theoretical Physics, Kiev, Ukraine
\item \Idef{org1262}Budker Institute for Nuclear Physics, Novosibirsk, Russia
\item \Idef{org1292}California Polytechnic State University, San Luis Obispo, California, United States
\item \Idef{org14939}Centre de Calcul de l'IN2P3, Villeurbanne, France
\item \Idef{org1197}Centro de Aplicaciones Tecnol\'{o}gicas y Desarrollo Nuclear (CEADEN), Havana, Cuba
\item \Idef{org1242}Centro de Investigaciones Energ\'{e}ticas Medioambientales y Tecnol\'{o}gicas (CIEMAT), Madrid, Spain
\item \Idef{org1244}Centro de Investigaci\'{o}n y de Estudios Avanzados (CINVESTAV), Mexico City and M\'{e}rida, Mexico
\item \Idef{org1335}Centro Fermi -- Centro Studi e Ricerche e Museo Storico della Fisica ``Enrico Fermi'', Rome, Italy
\item \Idef{org17347}Chicago State University, Chicago, United States
\item \Idef{org1288}Commissariat \`{a} l'Energie Atomique, IRFU, Saclay, France
\item \Idef{org1294}Departamento de F\'{\i}sica de Part\'{\i}culas and IGFAE, Universidad de Santiago de Compostela, Santiago de Compostela, Spain
\item \Idef{org1106}Department of Physics Aligarh Muslim University, Aligarh, India
\item \Idef{org1121}Department of Physics and Technology, University of Bergen, Bergen, Norway
\item \Idef{org1162}Department of Physics, Ohio State University, Columbus, Ohio, United States
\item \Idef{org1300}Department of Physics, Sejong University, Seoul, South Korea
\item \Idef{org1268}Department of Physics, University of Oslo, Oslo, Norway
\item \Idef{org1132}Dipartimento di Fisica dell'Universit\`{a} and Sezione INFN, Bologna, Italy
\item \Idef{org1270}Dipartimento di Fisica dell'Universit\`{a} and Sezione INFN, Padova, Italy
\item \Idef{org1315}Dipartimento di Fisica dell'Universit\`{a} and Sezione INFN, Trieste, Italy
\item \Idef{org1145}Dipartimento di Fisica dell'Universit\`{a} and Sezione INFN, Cagliari, Italy
\item \Idef{org1312}Dipartimento di Fisica dell'Universit\`{a} and Sezione INFN, Turin, Italy
\item \Idef{org1285}Dipartimento di Fisica dell'Universit\`{a} `La Sapienza' and Sezione INFN, Rome, Italy
\item \Idef{org1154}Dipartimento di Fisica e Astronomia dell'Universit\`{a} and Sezione INFN, Catania, Italy
\item \Idef{org1290}Dipartimento di Fisica `E.R.~Caianiello' dell'Universit\`{a} and Gruppo Collegato INFN, Salerno, Italy
\item \Idef{org1103}Dipartimento di Scienze e Innovazione Tecnologica dell'Universit\`{a} del Piemonte Orientale and Gruppo Collegato INFN, Alessandria, Italy
\item \Idef{org1114}Dipartimento Interateneo di Fisica `M.~Merlin' and Sezione INFN, Bari, Italy
\item \Idef{org1237}Division of Experimental High Energy Physics, University of Lund, Lund, Sweden
\item \Idef{org1192}European Organization for Nuclear Research (CERN), Geneva, Switzerland
\item \Idef{org1227}Fachhochschule K\"{o}ln, K\"{o}ln, Germany
\item \Idef{org1122}Faculty of Engineering, Bergen University College, Bergen, Norway
\item \Idef{org1136}Faculty of Mathematics, Physics and Informatics, Comenius University, Bratislava, Slovakia
\item \Idef{org1274}Faculty of Nuclear Sciences and Physical Engineering, Czech Technical University in Prague, Prague, Czech Republic
\item \Idef{org1229}Faculty of Science, P.J.~\v{S}af\'{a}rik University, Ko\v{s}ice, Slovakia
\item \Idef{org1184}Frankfurt Institute for Advanced Studies, Johann Wolfgang Goethe-Universit\"{a}t Frankfurt, Frankfurt, Germany
\item \Idef{org1215}Gangneung-Wonju National University, Gangneung, South Korea
\item \Idef{org1212}Helsinki Institute of Physics (HIP) and University of Jyv\"{a}skyl\"{a}, Jyv\"{a}skyl\"{a}, Finland
\item \Idef{org1203}Hiroshima University, Hiroshima, Japan
\item \Idef{org1329}Hua-Zhong Normal University, Wuhan, China
\item \Idef{org1254}Indian Institute of Technology, Mumbai, India
\item \Idef{org36378}Indian Institute of Technology Indore (IIT), Indore, India
\item \Idef{org1266}Institut de Physique Nucl\'{e}aire d'Orsay (IPNO), Universit\'{e} Paris-Sud, CNRS-IN2P3, Orsay, France
\item \Idef{org1277}Institute for High Energy Physics, Protvino, Russia
\item \Idef{org1249}Institute for Nuclear Research, Academy of Sciences, Moscow, Russia
\item \Idef{org1320}Nikhef, National Institute for Subatomic Physics and Institute for Subatomic Physics of Utrecht University, Utrecht, Netherlands
\item \Idef{org1250}Institute for Theoretical and Experimental Physics, Moscow, Russia
\item \Idef{org1230}Institute of Experimental Physics, Slovak Academy of Sciences, Ko\v{s}ice, Slovakia
\item \Idef{org1127}Institute of Physics, Bhubaneswar, India
\item \Idef{org1275}Institute of Physics, Academy of Sciences of the Czech Republic, Prague, Czech Republic
\item \Idef{org1139}Institute of Space Sciences (ISS), Bucharest, Romania
\item \Idef{org27399}Institut f\"{u}r Informatik, Johann Wolfgang Goethe-Universit\"{a}t Frankfurt, Frankfurt, Germany
\item \Idef{org1185}Institut f\"{u}r Kernphysik, Johann Wolfgang Goethe-Universit\"{a}t Frankfurt, Frankfurt, Germany
\item \Idef{org1177}Institut f\"{u}r Kernphysik, Technische Universit\"{a}t Darmstadt, Darmstadt, Germany
\item \Idef{org1256}Institut f\"{u}r Kernphysik, Westf\"{a}lische Wilhelms-Universit\"{a}t M\"{u}nster, M\"{u}nster, Germany
\item \Idef{org1246}Instituto de Ciencias Nucleares, Universidad Nacional Aut\'{o}noma de M\'{e}xico, Mexico City, Mexico
\item \Idef{org1247}Instituto de F\'{\i}sica, Universidad Nacional Aut\'{o}noma de M\'{e}xico, Mexico City, Mexico
\item \Idef{org23333}Institut of Theoretical Physics, University of Wroclaw
\item \Idef{org1308}Institut Pluridisciplinaire Hubert Curien (IPHC), Universit\'{e} de Strasbourg, CNRS-IN2P3, Strasbourg, France
\item \Idef{org1182}Joint Institute for Nuclear Research (JINR), Dubna, Russia
\item \Idef{org1143}KFKI Research Institute for Particle and Nuclear Physics, Hungarian Academy of Sciences, Budapest, Hungary
\item \Idef{org1199}Kirchhoff-Institut f\"{u}r Physik, Ruprecht-Karls-Universit\"{a}t Heidelberg, Heidelberg, Germany
\item \Idef{org20954}Korea Institute of Science and Technology Information, Daejeon, South Korea
\item \Idef{org1160}Laboratoire de Physique Corpusculaire (LPC), Clermont Universit\'{e}, Universit\'{e} Blaise Pascal, CNRS--IN2P3, Clermont-Ferrand, France
\item \Idef{org1194}Laboratoire de Physique Subatomique et de Cosmologie (LPSC), Universit\'{e} Joseph Fourier, CNRS-IN2P3, Institut Polytechnique de Grenoble, Grenoble, France
\item \Idef{org1187}Laboratori Nazionali di Frascati, INFN, Frascati, Italy
\item \Idef{org1232}Laboratori Nazionali di Legnaro, INFN, Legnaro, Italy
\item \Idef{org1125}Lawrence Berkeley National Laboratory, Berkeley, California, United States
\item \Idef{org1234}Lawrence Livermore National Laboratory, Livermore, California, United States
\item \Idef{org1251}Moscow Engineering Physics Institute, Moscow, Russia
\item \Idef{org1140}National Institute for Physics and Nuclear Engineering, Bucharest, Romania
\item \Idef{org1165}Niels Bohr Institute, University of Copenhagen, Copenhagen, Denmark
\item \Idef{org1109}Nikhef, National Institute for Subatomic Physics, Amsterdam, Netherlands
\item \Idef{org1283}Nuclear Physics Institute, Academy of Sciences of the Czech Republic, \v{R}e\v{z} u Prahy, Czech Republic
\item \Idef{org1264}Oak Ridge National Laboratory, Oak Ridge, Tennessee, United States
\item \Idef{org1189}Petersburg Nuclear Physics Institute, Gatchina, Russia
\item \Idef{org1170}Physics Department, Creighton University, Omaha, Nebraska, United States
\item \Idef{org1157}Physics Department, Panjab University, Chandigarh, India
\item \Idef{org1112}Physics Department, University of Athens, Athens, Greece
\item \Idef{org1152}Physics Department, University of Cape Town, iThemba LABS, Cape Town, South Africa
\item \Idef{org1209}Physics Department, University of Jammu, Jammu, India
\item \Idef{org1207}Physics Department, University of Rajasthan, Jaipur, India
\item \Idef{org1200}Physikalisches Institut, Ruprecht-Karls-Universit\"{a}t Heidelberg, Heidelberg, Germany
\item \Idef{org1325}Purdue University, West Lafayette, Indiana, United States
\item \Idef{org1281}Pusan National University, Pusan, South Korea
\item \Idef{org1176}Research Division and ExtreMe Matter Institute EMMI, GSI Helmholtzzentrum f\"ur Schwerionenforschung, Darmstadt, Germany
\item \Idef{org1334}Rudjer Bo\v{s}kovi\'{c} Institute, Zagreb, Croatia
\item \Idef{org1298}Russian Federal Nuclear Center (VNIIEF), Sarov, Russia
\item \Idef{org1252}Russian Research Centre Kurchatov Institute, Moscow, Russia
\item \Idef{org1224}Saha Institute of Nuclear Physics, Kolkata, India
\item \Idef{org1130}School of Physics and Astronomy, University of Birmingham, Birmingham, United Kingdom
\item \Idef{org1338}Secci\'{o}n F\'{\i}sica, Departamento de Ciencias, Pontificia Universidad Cat\'{o}lica del Per\'{u}, Lima, Peru
\item \Idef{org1316}Sezione INFN, Trieste, Italy
\item \Idef{org1271}Sezione INFN, Padova, Italy
\item \Idef{org1313}Sezione INFN, Turin, Italy
\item \Idef{org1286}Sezione INFN, Rome, Italy
\item \Idef{org1146}Sezione INFN, Cagliari, Italy
\item \Idef{org1133}Sezione INFN, Bologna, Italy
\item \Idef{org1115}Sezione INFN, Bari, Italy
\item \Idef{org1155}Sezione INFN, Catania, Italy
\item \Idef{org1322}Soltan Institute for Nuclear Studies, Warsaw, Poland
\item \Idef{org36377}Nuclear Physics Group, STFC Daresbury Laboratory, Daresbury, United Kingdom
\item \Idef{org1258}SUBATECH, Ecole des Mines de Nantes, Universit\'{e} de Nantes, CNRS-IN2P3, Nantes, France
\item \Idef{org1304}Technical University of Split FESB, Split, Croatia
\item \Idef{org1168}The Henryk Niewodniczanski Institute of Nuclear Physics, Polish Academy of Sciences, Cracow, Poland
\item \Idef{org17361}The University of Texas at Austin, Physics Department, Austin, TX, United States
\item \Idef{org1173}Universidad Aut\'{o}noma de Sinaloa, Culiac\'{a}n, Mexico
\item \Idef{org1296}Universidade de S\~{a}o Paulo (USP), S\~{a}o Paulo, Brazil
\item \Idef{org1149}Universidade Estadual de Campinas (UNICAMP), Campinas, Brazil
\item \Idef{org1239}Universit\'{e} de Lyon, Universit\'{e} Lyon 1, CNRS/IN2P3, IPN-Lyon, Villeurbanne, France
\item \Idef{org1205}University of Houston, Houston, Texas, United States
\item \Idef{org20371}University of Technology and Austrian Academy of Sciences, Vienna, Austria
\item \Idef{org1222}University of Tennessee, Knoxville, Tennessee, United States
\item \Idef{org1310}University of Tokyo, Tokyo, Japan
\item \Idef{org1318}University of Tsukuba, Tsukuba, Japan
\item \Idef{org21360}Eberhard Karls Universit\"{a}t T\"{u}bingen, T\"{u}bingen, Germany
\item \Idef{org1225}Variable Energy Cyclotron Centre, Kolkata, India
\item \Idef{org1306}V.~Fock Institute for Physics, St. Petersburg State University, St. Petersburg, Russia
\item \Idef{org1323}Warsaw University of Technology, Warsaw, Poland
\item \Idef{org1179}Wayne State University, Detroit, Michigan, United States
\item \Idef{org1260}Yale University, New Haven, Connecticut, United States
\item \Idef{org1332}Yerevan Physics Institute, Yerevan, Armenia
\item \Idef{org15649}Yildiz Technical University, Istanbul, Turkey
\item \Idef{org1301}Yonsei University, Seoul, South Korea
\item \Idef{org1327}Zentrum f\"{u}r Technologietransfer und Telekommunikation (ZTT), Fachhochschule Worms, Worms, Germany
\end{Authlist}
\endgroup

%
%
\end{document}